\documentclass[aps,pre,twocolumn,showpacs]{revtex4-1}
\usepackage{amsmath} 
\usepackage{amsfonts} 
\usepackage{graphicx} 

\begin{document}
\title{Nonequilibrium current-carrying steady states in the anisotropic $XY$ spin chain}
\author{Jarrett L. Lancaster}
\affiliation{Department of Nanoscience, Joint School of Nanoscience and Nanoengineering, The University of North Carolina at Greensboro, 2907 E. Gate City Boulevard, Greensboro, NC 27401} %
\date{\today}
\begin{abstract}
Out-of-equilibrium behavior is explored in the one-dimensional anisotropic $XY$ model. Initially preparing the system in the isotropic $XX$ model with a linearly varying magnetic field to create a domain-wall magnetization profile, dynamics is generated by rapidly changing the exchange interaction anisotropy and external magnetic field. Relaxation to a nonequilibrium steady state is studied analytically at the critical transverse Ising point, where correlation functions may be computed in closed form. For arbitrary values of anisotropy and external field, an effective generalized Gibbs' ensemble is shown to accurately describe observables in the long-time limit. Additionally, we find spatial oscillations in the exponentially decaying, transverse spin-spin correlation functions with wavelength set by the magnetization jump across the initial domain wall. This wavelength depends only weakly on anisotropy and magnetic field in contrast to the current, which is highly dependent on these parameters.
\end{abstract}


\maketitle

\section{Introduction}
\label{sec:intro}
Recent experiments with cold atoms have resulted in a deluge of theoretical work in the area of quantum dynamics of closed, many-body systems \cite{Silva2011,Cazalilla2011,Eisert2015}. The ability to engineer faithful representations of low-dimensional models with tunable interactions has ushered in the age of quantum simulators \cite{Feynman1982}. By carefully exploiting Feshbach resonances \cite{Chin2010} and precisely tuning optical lattices one may ``dial up'' a large variety of model systems, such as hard-core bosons \cite{Bloch2015} or quantum spin chains \cite{Senko2015} and observe the isolated unitary dynamics of large numbers of strongly interacting particles.

This recent explosion of experiments probing strongly correlated dynamics leaves theorists with the task of characterizing the dynamics of observables in systems driven far from equilibrium. Definitive experimental evidence of interacting systems that do not approach thermal equilibrium within significant timescales \cite{Weiss2006} has provided a valuable experimental component to the theoretical investigations of integrable models with dynamics tightly constrained by the existence of a conserved quantity for each degree of freedom. Indeed, one of the fascinating results of the past decade has been the establishment of the generalized Gibbs' ensemble (GGE) framework for obtaining long-time averages of observables in integrable systems \cite{Rigol2006,Rigol2007,Rigol2008,Cazalilla2006}.  For weak integrability-breaking perturbations, these generalized Gibbs' ensemble averages often correspond to long-lived ``pre-thermalization plateaus'' with thermalization occurring on extremely large timescales \cite{Kollar2011,Marcuzzi2013}. Careful measurements of multi-point correlation functions during the relaxation of a one-dimensional Bose gas have established its approach to a state precisely captured by a statistical description for timescales on the order on milliseconds \cite{Langen2015}. Additionally, a generalization of the fluctuation-dissipation allows for the formal definition of a frequency-dependent (mode-dependent) effective temperature in integrable systems far from equilibrium \cite{Bortolin2015}.


Treatment of interactions away from equilibrium is generally challenging, especially for spatially inhomogeneous initial states. The time-dependent density matrix renormalization group (tDMRG) provides a popular numerical approach for studying the evolution of interacting systems and has been applied to the $XXZ$ spin chain for initial states with domain-wall magnetization profiles \cite{Schollwock2005}. Application of low-energy, continuum methods such as bosonization to out-of-equilibrium dynamics in the $XXZ$ spin chain has provided predictions \cite{Lancaster2010a,Foster2010} which compare quite well to numerical treatments \cite{Foster2011,Sabetta2013}, at least for weak interactions. Good agreement between continuum bosonization predictions and tDMRG numerical results has also been found for spatially inhomogeneous initial states in spin Heisenberg chains with next-nearest neighbor interactions \cite{Bravo2013}. Understanding how the low-energy, continuum description is modified far from equilibrium is also an active area of research, with significant efforts being made to investigate the roles of operators which are irrelevant in equilibrium but strongly affect the physics away from equilibrium \cite{Mitra2011,Mitra2012,Tavora2013}. The majority of recent theoretical efforts have focused on the limiting cases for which the exchange couplings are equal, $J_{x} = J_{y}\,$ ($XX$, $XXZ$ models), or in which only one coupling is nonzero (the quantum Ising model) \cite{Cazalilla2012}. Experimental realizations of inhomogeneous states of weakly interacting, two-level pseudospin systems have been presented previously \cite{Weld2009}, making experimental investigations of the non-equilibrium dynamics arising from spatially inhomogeneous initial states feasible.

The $XY$ spin chain, first investigated by Lieb, Schultz and Mattis \cite{LSM1961} and later generalized to include an external magnetic field \cite{Barouch1,Barouch2,Pfeuty1970} is a simply stated system with a rich phase diagram. Mapping to free fermions through a Jordan-Wigner transformation, interactions between quasiparticles correspond to interactions between nearest-neighboring $z$-components of spin. Consequently, the $XY$ chain and its immediate generalizations provide a fruitful playground for theoretical investigations of one-dimensional quantum dynamics far from equilibrium. In particular, magnetization dynamics arising from simple ``quantum quenches,'' or rapid changes in system parameters, was first investigated decades ago \cite{Barouch2} by computing observables after suddenly changing the strength of an external magnetic field. Recent efforts have concentrated on current-carrying steady states in the isotropic $XX$ limit \cite{Antal1997,Antal1998}, evolution of initial domain wall magnetization profiles \cite{Antal1999,Lancaster2010a,Karevski2002}, and behavior of correlation functions in the Ising limit after rapid quenches of the magnetic field \cite{Rossini2011,Caneva2011}. Investigations of so-called geometric quenches have also yielded exact results for the $XX\,$ limit and numerical results for short time dynamics in the $XXZ\,$ chain \cite{Alba2014}. A number of previous works have also focused on quench dynamics arising from slow changes in the magnetic field \cite{Mukherjee2007} and anisotropy parameter \cite{Mukherjee2008} in the $XY$ chain. Recent results have been obtained regarding sufficient conditions for the appearance of Kibble-Zurek scaling (KZS) which depend on details of the initial quasi-particle distribution function \cite{Deng2011}. Field theoretic methods \cite{Cardy2006} have been applied to investigate the general features of correlation functions after a quench, with some results for inhomogeneous initial states also available \cite{Cardy2008,Calabrese2008}. Analytic expressions have also been obtained for work statistics after sudden anisotropy quenches \cite{Bayocboc2015} for zero and nonzero temperature initial states.

As a special case of the $XY$ spin chain, the one-dimensional quantum Ising model has received a significant amount of attention due to its simplicity, and there are indications that its nonequilibrium dynamics can be probed by employing circuit quantum electrodynamics (QED) \cite{Viehmann2013}. Dynamical quantum phase transitions have been theoretically investigated in both the bare transverse-Ising (TI) chain \cite{Heyl2013} as well as the extended axial next-nearest neighbor Ising (ANNNI) chain \cite{Kriel2014}. Additional evidence of the rich theoretical structure was provided by the discovery of underlying $E_{8}\,$ group structure in a perturbed transverse Ising chain \cite{Zamolodchikov1989,Mussardo} through a formal mapping to a two-dimensional classical Ising field theory. Recently, experiments have led to striking, if low-resolution, observations \cite{Coldea2010} through neutron scattering which are consistent with this emergent group structure. Ising-like models with long-range interactions have also been recently created experimentally in reduced dimensions \cite{Richerme2014,Jurcevic2015} using cold atomic gases.

Much of the theoretical work on spin-chain dynamics has involved a sudden quench in the system's transverse field with homogeneous initial states \cite{Calabrese2011,Foini2011}. In this paper, we investigate a sudden quench in the strength of anisotropy and external magnetic field while also taking the initial state to describe an inhomogeneous, domain-wall magnetization profile. While some basic features of domain walls in the Ising model have been previously considered \cite{Karevski2002,Subrahmanyam2003}, the focus of this paper is to thoroughly investigate the time-dependent magnetization, emergent spin current and long-time limit of the two-point, equal-time correlation functions. We will present analytic results for these observables after quenching to the critical transverse-Ising limit and demonstrate the existence of an effective steady state for the central region of the system after the domain wall has effectively flattened out over a finite region of space.

The $XY$ model is described by the following Hamiltonian,
\begin{eqnarray}
\hat{H} & = & -\sum_{j}\left[J_{x}\hat{S}_{j}^{x}\hat{S}_{j+1}^{x} + J_{y}\hat{S}^{y}_{j}\hat{S}^{y}_{j+1}\right] + \sum_{j}h_{j}\hat{S}_{j}^{z}.\label{eq:h1}
\end{eqnarray} 
Extending some recent results for the isotropic $XX$ model \cite{Sabetta2013}, we are able to extract long-time behavior of observables away from the two critical lines and describe the non-equilibrium steady state for $0 \leq  \left|\frac{h}{J}\right| < 1\,$ with arbitrary $J_{x}, J_{y}>0$. The particularly simple isotropic limit $J_{x} = J_{y} \equiv J\,$ is known as the isotropic $XY\,$ model, or $XX$ model. However, for any nonzero value of $\gamma \equiv \frac{J_{x}-J_{y}}{J_{x}+J_{y}}\,$ and constant magnetic field $h_{j} \rightarrow h$, the model can be diagonalized, as the system maps to free fermions. In what follows we set the lattice spacing $a = 1$. Interactions may be included by adding the following term to Eq.~(\ref{eq:h1})
\begin{equation}
\hat{V} = J_{z}\sum_{j}\hat{S}_{j}^{z}\hat{S}_{j+1}^{z}.\label{eq:v}
\end{equation}
For $\gamma = 0$, the resulting interacting system described by the sum of Eqs.~(\ref{eq:h1}) and (\ref{eq:v}) is known as the $XXZ$ chain, which is gapless for $\frac{J_{z}}{J}<1$. In this paper we specialize to the non-interacting system $J_{z} = 0$.

Creating domain-wall magnetization profiles as initial states is an efficient method of generating non-equilibrium dynamics. For the isotropic $XX$ model, the dynamics is quite easily understood as ballistic expansion of a gas of free fermions \cite{Rigol2004,Rigol2006,Lancaster2010a}. The more complex $XXZ$ chain involves interactions between the fermions, but for weak interactions the basic picture appears largely unchanged. Bosonization, a low-energy approach for describing one-dimensional interacting systems, works surprisingly well in this far-from-equilibrium setting \cite{Lancaster2010a,Sabetta2013}. At long times, after the magnetization in a central region of the system has essentially relaxed to its equilibrium value, one finds that spatial oscillations are superimposed on the algebraic decay present in the transverse-spin correlation function, 
\begin{equation}
\lim_{t\rightarrow\infty}\left\langle \hat{S}^{x}_{j}(t)\hat{S}^{x}_{j+n}(t)\right\rangle = \cos\left(\frac{2\pi n}{\lambda}\right)\left\langle \hat{S}_{j}^{x}\hat{S}_{j+n}^{x}\right\rangle_{\mbox{\scriptsize g.s.}},
\end{equation}
with $\lambda = \frac{2}{m_{0}}\,$ \cite{Lancaster2010a} and $m_{0}\,$ the magnitude of magnetization far from the domain. Here the subscript ``g.s.'' refers to the expression one would obtain by considering the equilibrium behavior of the system at $T=0\,$ in which the system is in its ground state. In the language of hard-core bosons, this twist is related to quasi-condensation at $k_{\pm} = \pm\frac{\pi}{2}\,$ during rapid expansion, which has recently been observed experimentally \cite{Bloch2015}. Representing the system in terms of free fermions through the Jordan-Wigner transformation, both effects may be understood as emerging from an underlying particle current. In the isotropic $XY$ model, this current emerges according to a conservation law requiring a local change in magnetization to be accompanied by a spin current,
\begin{eqnarray}
\partial_{t}\hat{S}_{j}^{z} & = & -\left[\hat{\mathcal{J}}_{j+1}-\hat{\mathcal{J}}_{j}\right],\\
\hat{\mathcal{J}}_{j} & = & \hat{S}_{j+1}^{+}\hat{S}_{j}^{-} - \hat{S}_{j}^{+}\hat{S}_{j+1}^{-}\;\;\;\;\;(XX\mbox{-model})\label{eq:cur}.
\end{eqnarray}
Antal and collaborators studied observables in the $XX$ chain with constant (nonzero) currents through the introduction of a chemical potential-like term involving the total current operator to the Hamiltonian \cite{Antal1997,Antal1998}. Interestingly, the long-time limit of domain-wall states in the gapless phase of the $XXZ\,$ model leads to a non-equilibrium steady state characterized by ``twisted'' single-particle Green's functions \cite{Lancaster2010a,Sabetta2013},
\begin{equation}
\lim_{t\rightarrow \infty}\left\langle c_{j}^{\dagger}(t)c_{j+n}(t)\right\rangle = e^{in\phi}\left\langle c_{j}^{\dagger}c_{j+n}\right\rangle_{\mbox{\scriptsize g.s.}},
\end{equation}
where $\phi \propto m_{0}$, and $m_{0}\,$ is the ``height'' of the domain wall. In the noninteracting $XX$ model, $\left\langle \hat{\mathcal{J}}^{z}_{j}\right\rangle = \frac{J}{\pi}\sin\left(\pi m_{0}\right) = \frac{J}{\pi}\sin\phi$, so that the wavelength of spatial oscillations is directly related to the spin current. The presence of weak interactions has been shown to not strongly modify this picture in an essential way \cite{Lancaster2010a,Sabetta2013}, though strong interactions may greatly influence the details of the oscillations and decay of correlations \cite{Lancaster2010b}. 

The main goal of this paper is to explore how anisotropy between the $x$ and $y$ components of nearest-neighbor exchange interactions affects the relationship between initial domain wall height, long-time current and spatial oscillations in $\left\langle\hat{S}_{j}^{x}(t)\hat{S}_{j+n}^{x}(t)\right\rangle$. For $\gamma \neq 0$, the system no longer conserves total magnetization,
\begin{equation}
\left[ \sum_{j}\hat{S}_{j}^{z} , \hat{H}_{\gamma\neq 0}\right] \neq 0,
\end{equation}
so that the spin current in Eq.~(\ref{eq:cur}) no longer corresponds to a local conservation law. However, this spin current does possess nonzero overlap with the actual fermion current obtained after mapping the system to noninteracting fermions through a Bogoliubov rotation.  For this reason, a piece of current survives in the long-time limit despite the actual operator not being totally conserved. 

This paper is organized as follows. In Sec.~\ref{sec:xy} we review some relevant background information on the ground-state properties of $XX$ and $XY$ spin chains and provide a brief summary of domain-wall dynamics in $XX\,$ spin chains. Section~\ref{sec:domaincalc} contains analytic calculations of observables when a domain-wall spin configuration in the $XX$-model is subjected to a rapid change in anisotropy and external magnetic field so that the time evolution takes place with respect to the transverse Ising model on the critical line $h = J\,$ with $\gamma = 1$. Emphasis is placed on the long-time average behavior of the magnetization, current and spin-spin correlation functions. In Sec.~\ref{sec:ness}, an effective nonequilibrium steady state is shown to accurately describe the long-time limit of observables within the central region of the chain. Furthermore, it is shown that observables in this non-equilibrium steady state are easily calculated away from the critical lines and for more general domain walls with arbitrary jumps in magnetization.   Results are summarized and discussed in Sec.~\ref{sec:concl}.

\section{Model}
\label{sec:xy}
To formulate the spreading domain-wall problem as a single ``quantum quench,'' we consider the following protocol: The system is initialized as the ground state of an infinite $XX$ spin chain with a spatially varying magnetic field,
\begin{equation}
\hat{H}_{0} = -J\sum_{j}\left[\hat{S}_{j}^{x}\hat{S}_{j+1}^{x} + \hat{S}_{j}^{y}\hat{S}_{j+1}^{y}\right] + \sum_{j}h_{j}\hat{S}_{j}^{z},\label{eq:hinit}
\end{equation}
where the field $h_{j}\,$ acts as a spatially varying chemical potential for the quasiparticles of the system (see below). At $t = 0$, the field $h_{j}\,$ is abruptly switched to a constant value at all lattice sites $h$, and nonzero anisotropy $\gamma\,$ between the nearest-neighbor spin interactions is switched on, resulting in time evolution being generated by the anisotropic $XY\,$ model,
\begin{eqnarray}
\hat{H}_{f} & = &  -J\sum_{j}\left[(1+\gamma)\hat{S}_{j}^{x}\hat{S}_{j+1}^{x} + (1-\gamma) \hat{S}_{j}^{y}\hat{S}_{j+1}^{y} \right]\nonumber\\
& + &  h\sum_{j}\hat{S}_{j}^{z}.\label{eq:xy}
\end{eqnarray}
The system is generally gapped for nonzero $\gamma\,$ and $h\,$ except along the critical line $\gamma =0\,$ with $\left|\frac{h}{J}\right|<1\,$ and in the critical transverse-Ising limit ($\gamma = \frac{h}{J} = 1$). Ground state observables and correlation functions in the presence of arbitrary anisotropy and magnetic fields were extensively studied by Barouch and collaborators \cite{Barouch1,Barouch2,Barouch3,Barouch4}. In this paper we will investigate the non-equilibrium dynamics within the larger $(\gamma,h/J)\,$ phase diagram, specifically in the range $0 \leq \gamma,\frac{h}{J} < 1$. The gapless cases provide instances in which the observable calculations may be performed entirely analytically.
\subsection{Ground state properties of homogeneous $XY\,$ chains}
It is useful to review the basic ground-state observables in the $XY$ spin chain as a point of comparison for the non-equilibrium results to be presented. Equation~(\ref{eq:xy}) is diagonalized in terms of non-interacting fermions by a Jordan-Wigner transformation \cite{LSM1961}
\begin{eqnarray}
\hat{S}^{+}_{j} & = & c_{j}^{\dagger}\exp\left[i\pi\sum_{n=1}^{j-1}c_{n}^{\dagger}c_{n}\right],\label{eq:jw1}\\
\hat{S}^{z}_{j} & = & c_{j}^{\dagger}c_{j}-\frac{1}{2}.\label{eq:jw2}
\end{eqnarray}
After applying Eqs.~(\ref{eq:jw1}) and (\ref{eq:jw2}), Eq.~(\ref{eq:xy}) becomes
\begin{eqnarray}
\hat{H}_{f} & = & -\frac{J}{2}\sum_{j}\left[c_{j}^{\dagger}c_{j+1} +  c_{j+1}^{\dagger}c_{j} + \gamma c_{j}^{\dagger}c_{j+1}^{\dagger} + \gamma c_{j+1}c_{j}\right]\nonumber\\
& + & h\sum_{j}c_{j}^{\dagger}c_{j}.
\end{eqnarray} 
This quadratic Hamiltonian may be brought into diagonal form
\begin{equation}
\hat{H}_{f} = \sum_{k}\epsilon_{k}\eta_{k}^{\dagger}\eta_{k},
\end{equation}
with
\begin{equation}
\epsilon_{k} = -J\mbox{sgn}\left(\cos k - \frac{h}{J}\right)\sqrt{\left(\cos k - \frac{h}{J}\right)^{2} + \gamma^{2}\sin^{2}k},\label{eq:xydisp}
\end{equation}
where the quasiparticles are related to the original momentum-space operators according to a Bogoliubov transformation,
\begin{equation}
\left(\begin{array}{c} c_{k}\\c_{-k}^{\dagger}\end{array}\right) = \left(\begin{array}{cc} \cos\frac{\theta_{k}}{2} & -i\sin\frac{\theta_{k}}{2} \\ -i\sin\frac{\theta_{k}}{2} & \cos\frac{\theta_{k}}{2}\end{array}\right)\left(\begin{array}{c} \eta_{k}\\ \eta_{-k}^{\dagger}\end{array}\right),\label{eq:bog}
\end{equation}
with 
\begin{eqnarray}
\cos\theta_{k} & = & \frac{\left|\cos k - \frac{h}{J}\right|}{\sqrt{\left(\cos k - \frac{h}{J}\right)^{2}+\gamma^{2}\sin^{2}k}},\label{eq:sink}\\
\sin\theta_{k} & = & \frac{\mbox{sgn}\left(\cos k - \frac{h}{J}\right)\gamma \sin k}{\sqrt{\left(\cos k - \frac{h}{J}\right)^{2}+\gamma^{2}\sin^{2}k}},\label{eq:cosk}\\
c_{j} & = & \frac{1}{\sqrt{N}}\sum_{k}e^{ikj}c_{j}.\label{eq:momspace}
\end{eqnarray}
We are interested in the thermodynamic limit $N\rightarrow\infty\,$ in which the sum in Eq.~(\ref{eq:momspace}) may be converted to a continuous integral. The ground state is then occupied by all negative-energy quasiparticles
\begin{equation}
\left|\Psi_{0}\right\rangle = \prod_{\epsilon_{k}<0}\eta_{k}^{\dagger}\left|0\right\rangle,
\end{equation}
where the vacuum state obeys $\eta_{k}\left|0\right\rangle = 0\,$ for all $k$. This convention differs from original treatment \cite{LSM1961} in which particle-hole symmetry is employed to simplify the algebra by taking all energy eigenvalues as positive and reinterpreting the ground state as a vacuum for all excitations. In the context of inhomogeneous quench dynamics, keeping track of the signs within the present formulation turns out to be a small price to pay for a physically transparent picture. Spin-spin correlation functions at equal times are obtained in terms of Pfaffians, as discussed in Appendix~\ref{sec:pfaffian}. All observables can be written in terms of the basic contractions of Majorana operators 
\begin{eqnarray}
A_{j} & = & c_{j}^{\dagger} + c_{j},\label{eq:Aj}\\
B_{j} & = & c_{j}^{\dagger} - c_{j}.\label{eq:Bj}
\end{eqnarray}
For example,
\begin{eqnarray}
\left\langle \hat{S}_{j}^{z}(t)\right\rangle & = &  \left\langle c_{j}^{\dagger}(t)c_{j}(t)\right\rangle -\frac{1}{2},\\
& = & \frac{1}{2} \left\langle B_{j}(t)A_{j}(t)\right\rangle,
\end{eqnarray}
\begin{eqnarray}
& & \left\langle \hat{\mathcal{J}}_{j}^{z}(t)\right\rangle = J\mbox{Im}\left\langle c_{j}^{\dagger}(t)c_{j+1}(t)\right\rangle,\label{eq:cur1}\\
&  & = \frac{J}{4}\mbox{Im}\left\langle \left(B_{j}(t)+A_{j}(t)\right)\left(A_{j+1}(t)-B_{j+1}(t)\right)\right\rangle.
\end{eqnarray}
In the isotropic $XX$ limit with $h=0$, one finds \cite{LSM1961,Barouch2}
\begin{eqnarray}
\left\langle \hat{S}_{j}^{z}\hat{S}_{j+n}^{z}\right\rangle & = & \left\{\begin{array}{cc}\displaystyle -\frac{\sin^{2}\left(\frac{\pi n}{2}\right)}{\pi^{2}n^{2}} & (|n|>0)\\ \displaystyle \frac{1}{4} & (n=0)\end{array}\right.\\
\left\langle \hat{S}_{j}^{x}\hat{S}_{j+n}^{x}\right\rangle & = & \frac{e^{\frac{1}{2}}2^{\frac{2}{3}}}{\mathcal{A}^{6}\sqrt{n}}\left(1 + \frac{(-1)^{n}}{8n^{2}} + \cdots \right).
\end{eqnarray}
In this isotropic case $\left\langle \hat{S}_{j}^{y}\hat{S}_{j+n}^{y}\right\rangle =\left\langle \hat{S}_{j}^{x}\hat{S}_{j+n}^{x}\right\rangle\,$ by rotational invariance. In the transverse-Ising limit, $\gamma = \frac{h}{J} = 1$, one finds
\begin{eqnarray}
\left\langle \hat{S}_{j}^{z}\hat{S}_{j+n}^{z}\right\rangle & = & \frac{1}{\pi^{2}}\cdot\frac{1}{4n^{2}-1},\label{eq:ztieq}\\
\left\langle \hat{S}_{j}^{x}\hat{S}_{j+n}^{x}\right\rangle & = & \frac{e^{\frac{1}{4}}2^{\frac{1}{12}}}{4\mathcal{A}^{3}n^{\frac{1}{4}}}\left(1 - \frac{1}{64 n^{2}} + \cdots \right),\label{eq:xtieq}\\
\left\langle \hat{S}_{j}^{y}\hat{S}_{j+n}^{y}\right\rangle & = & -\frac{e^{\frac{1}{4}}2^{\frac{1}{12}}}{16\mathcal{A}^{3}n^{\frac{9}{4}}}\left(1 + \frac{15}{64 n^{2}} + \cdots \right),\label{eq:ytieq}
\end{eqnarray}
where $\mathcal{A} = 1.2824...\,$ is the Glaisher-Kinkelin constant. While power-law decay is observed for $\gamma = 0\,$ and along the line $h = J\,$ in the $\gamma-h\,$ plane, correlations generically decay exponentially and the spectrum is gapped everywhere except in the $XX\,$ and critical transverse-Ising limits.

Our main focus in the remainder of this paper is on investigating the nonequilibrium behavior of these correlation functions $\mathcal{C}_{n}^{\nu\nu}(j,t) \equiv \left\langle \hat{S}_{j}^{\nu}(t)\hat{S}_{j+n}^{\nu}(t)\right\rangle\,$ with $\nu = x,y,z$, as well as the simpler observables such as local magnetization and spin current. Expressions for the correlation functions $\mathcal{C}_{n}^{\nu\nu}(j,t)\,$ away from equilibrium are given in Appendix~\ref{sec:pfaffian}. Compared to $\mathcal{C}^{xx}_{n}\,$ and $\mathcal{C}^{yy}_{n}$, the form for $\mathcal{C}_{n}^{zz}(j,t)\,$ is rather simple,
\begin{equation}
\mathcal{C}_{n}^{zz}(j,t) = \frac{1}{4}\left\langle B_{j}(t)A_{j}(t)B_{j+n}(t)A_{j+n}(t)\right\rangle.\label{eq:czz}
\end{equation}
We explicitly obtain the time evolution of these observables for $\gamma = \frac{h}{J} = 1\,$ in Sec.~\ref{sec:domaincalc} to determine for which correlation functions the power-law correlations depicted in Eqs.~(\ref{eq:ztieq})-(\ref{eq:ytieq}) survive in the nonequilibrium setting. The behavior in other regions of the $\gamma-h\,$ plane is discussed using a generalized Gibbs' ensemble description of the central region at long times in Sec.~\ref{sec:ness}.

\subsection{Creating domain-wall magnetization profiles}
After applying the Jordan-Wigner transformation, the initial Hamiltonian Eq.~(\ref{eq:hinit}) becomes
\begin{equation}
\hat{H}_{0} = -\frac{J}{2}\sum_{j}\left[c_{j}^{\dagger}c_{j+1} + c_{j+1}^{\dagger}c_{j}\right] + \sum_{j}h_{j}c_{j}^{\dagger}c_{j}.
\end{equation}
Let us specialize to the case of a linearly varying magnetic field, $h_{j} = \mathcal{F}j$. This Hamiltonian maps to the single-particle Wannier-Stark problem \cite{Eisler2009} and may be diagonalized by employing a linear transformation,
\begin{equation}
\hat{H}_{0} = \sum_{m}\epsilon^{(0)}_{m}\beta_{m}^{\dagger}\beta_{m},
\end{equation}
with
\begin{equation}
\beta_{m} = \sum_{j}J_{j-m}(\alpha)c_{j},
\end{equation}
where $\alpha \equiv \frac{J}{\mathcal{F}}$, and $J_{n}(x)\,$ is the Bessel function of the first kind of order $n$. The ground state is populated by all single-particle states of negative energy,
\begin{equation}
\left|\Psi_{0}\right\rangle = \prod_{m<0}\beta_{m}^{\dagger}\left|0\right\rangle,
\end{equation}
with $\epsilon^{(0)}_{m} = m\mathcal{F}$. Far to the left (right) of the origin a very large positive (negative) magnetic field acts as a strong chemical potential creating a region of uniform positive (negative) magnetization. In terms of the Jordan-Wigner fermions, the left half of the system is filled with particles with unit $c$-fermion occupation number, $\left\langle \hat{n}_{j} \right\rangle = 1\,$ for $j \ll 0$ while the right half contains no particles, $\left\langle \hat{n}_{j}\right\rangle = 0$. Explicitly, the magnetization profile is
\begin{equation}
\left\langle \hat{S}_{j}^{z}\right\rangle = \left\langle \hat{n}_{j}\right\rangle - \frac{1}{2} = -\frac{1}{2} + \sum_{m>0}J_{j+m}^{2}(\alpha),\label{eq:szwall}
\end{equation}
which corresponds to a central spin gradient of width $\sim \alpha$, separating the regions of uniform maximal polarization. The dynamics arising from this state after abruptly switching $\mathcal{F} \rightarrow 0$ have previously been considered with the time-dependent magnetization \cite{Antal1999} and transverse correlations \cite{Lancaster2010a} computed explicitly. One finds the long-time behavior independent of $\alpha$, and for simplicity we consider the limit $\alpha \rightarrow 0$, corresponding to the sharp domain wall constructed from two semi-infinite, uniform regions of oppositely-polarized spins. Effectively, this allows us to write the initial state as
\begin{equation}
\left|\Psi_{0}\right\rangle = \prod_{j<0}c_{j}^{\dagger}\left|0\right\rangle.\label{eq:initstate}
\end{equation}
\begin{figure}
\begin{center}
\includegraphics[totalheight=6.0cm]{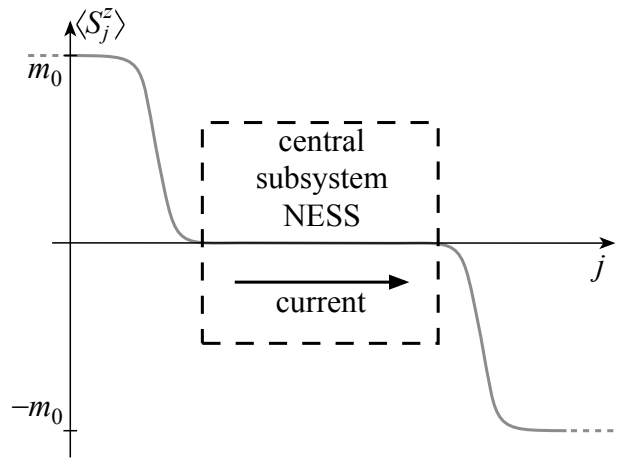}
\caption{Domain-wall broadening in the isotropic $XY$ model. As a domain-wall magnetization profile spreads, the central region relaxes to a non-equilibrium steady state. The presence of the polarized edges drives a current through the central region. This situation may be viewed as expansion of a quantum gas of interacting ($XXZ$ chain) or non-interacting ($XY$ chain) fermions in one dimension.}
\label{fig:nessfigj}
\end{center}
\end{figure}
When time evolution is generated by the $XX$ model, a long-time steady state emerges \cite{Antal1999,Lancaster2010a,Sabetta2013}. This steady state is characterized by a spin current, which is driven through the central region as the domain profile spreads, transferring net magnetization from the left side of the wall to the right side. The value of the current is set by the initial jump in magnetization between the far right and far left sides of the system and contains the extent of the central subsystem's ``memory'' of the initial state. Denoting the magnetization on the far left (right) side of the system by $+m_{0}\,$ ($-m_{0}$), we use the term ``domain-wall height'' to refer to $m_{0}$. That the long-time limit of the nonequilibrium behavior can be entirely described using only the knowledge of this magnetization jump has been demonstrated by Sabetta and Misguich \cite{Sabetta2013} in their explicit construction of the nonequilibrium steady state describing the central subsystem at long times. This region where the magnetization has essentially relaxed, as depicted in Fig.~\ref{fig:nessfigj}, is accurately described by a type of generalized Gibbs' ensemble which depends only on the initial magnetization jump $m_{0}$. While the initial states corresponding to domain walls with $m_{0} < \frac{1}{2}\,$ are somewhat more complicated than the state in Eq.~(\ref{eq:initstate}), we will address these more general domain walls by extending the results of Ref.~\onlinecite{Sabetta2013} to time evolution in the anisotropic $XY$-model in Sec.~\ref{sec:ness}.

\section{Domain wall dynamics in anisotropic $XY$-model at critical point}
\label{sec:domaincalc}
\subsection{$XY$-model dynamics}
The time evolution of the noninteracting quasiparticles is simple, with $\eta_{k}(t) = e^{-i\epsilon_{k}t}\eta_{k}$. When combined with Eq.~(\ref{eq:bog}), one may obtain the time dependence of the original $c$-fermions in terms of the initial operators,
\begin{equation}
\left(\begin{array}{c} c_{k}(t)\\ c_{-k}^{\dagger}(t)\end{array}\right) = \left(\begin{array}{cc} f_{k}(t) & g_{k}(t) \\ -g_{k}(t) & f_{k}^{*}(t)\end{array}\right)\left(\begin{array}{c} c_{k} \\c_{-k}^{\dagger}\end{array}\right),\label{eq:xyt}
\end{equation}
where
\begin{eqnarray}
f_{k}(t) & = & \cos(\epsilon_{k}t) - i\cos\theta_{k}\sin(\epsilon_{k}t),\label{eq:ft}\\
g_{k}(t) & = & \sin\theta_{k}\sin(\epsilon_{k}t).\label{eq:gt}
\end{eqnarray}
To obtain time-dependent operators in position space, we perform a final Fourier transform,
\begin{eqnarray}
c_{j}(t) & = & \frac{1}{\sqrt{N}}\sum_{k}e^{ikj}c_{k}(t),\\
& = & \sum_{n}\left[F_{j-n}(t)c_{n} + G_{j-n}(t)c_{n}^{\dagger}\right]\label{eq:cjt},
\end{eqnarray}
where
\begin{eqnarray}
F_{m}(t) & = & \int_{-\pi}^{\pi}\frac{dk}{2\pi}\left[\cos\left(\epsilon_{k}t\right)e^{ikm}\right.\nonumber\\
& - & \left. ie^{ikm}\cos\theta_{k}\sin(\epsilon_{k}t)\right],\label{eq:Fmt}\\
G_{m}(t) & = & \int_{-\pi}^{\pi}\frac{dk}{2\pi}e^{ikm}\left[\sin\theta_{k}\sin(\epsilon_{k}t)\right]\label{eq:Gmt}.
\end{eqnarray}
\subsection{Analytic solution for the TI critical point}
Equations~(\ref{eq:Fmt}) and (\ref{eq:Gmt}) may be integrated numerically for arbitrary values of $h\,$ and $\gamma$. However, in addition to the isotropic $\gamma =0\,$ line on the $(\gamma, \frac{h}{J})\,$ phase diagram where these expressions reduce to Bessel functions, the $XY\,$ model possesses another special limit in which closed-form expressions result. Namely, for $\frac{h}{J} = \gamma = 1$, the system is gapless and Eqs.~(\ref{eq:sink}) and (\ref{eq:cosk}) simplify to
\begin{eqnarray}
\cos\theta_{k} & = & \left|\sin\frac{k}{2}\right|,\\
\sin\theta_{k} & = & -\mbox{sgn}(k)\cos\frac{k}{2}.
\end{eqnarray}
Performing the integrations in Eqs.~(\ref{eq:Fmt}) and (\ref{eq:Gmt}) we obtain
\begin{eqnarray}
F_{m}(t) & = & J_{2m}(2Jt) - \frac{i}{2}\left[J_{2m+1}(2Jt)\right.\nonumber\\
& - &\left. J_{2m-1}(2Jt)\right]\label{eq:Ft},\\
G_{m}(t) & = & \frac{i}{2}\left[J_{2m+1}(2Jt) + J_{2m-1}(2Jt)\right]\label{eq:Gt},
\end{eqnarray}
where $J_{l}(z)\,$ is a Bessel function of the first kind of order $l$. Using Eqs.~(\ref{eq:Ft})-(\ref{eq:Gt}) and the initial state given in Eq.~(\ref{eq:initstate}), we may evaluate contractions of Eqs.~(\ref{eq:Aj})-(\ref{eq:Bj}), obtaining
\begin{widetext}
\begin{eqnarray}
\left\langle B_{j}(t)A_{j}(t)\right\rangle & = & J_{2j}J_{2j+2n} - J_{2j+1}J_{2j+2n-1}- \sum_{m=1}^{2j+n}\left[J_{2j-2m}J_{2j+2n-2m}- J_{2j-2m+1}J_{2j+2n-2m-1}\right],\label{eq:bat}
\end{eqnarray}
\begin{eqnarray}
\left\langle A_{j}(t)A_{j+n} (t)\right\rangle  & = &    i\left(J_{2j}J_{2j+2n-1}-J_{2j-1}J_{2j+2n}\right) + i\sum_{m=1}^{2j+n}\left[-J_{-2j+2m+1}J_{-2j-2n+2m}+ J_{-2j+2m}J_{-2j-2n+2m+1}\right]\nonumber \\
& + &  i\sum_{m>0}\left[J_{2j+2m}\left(J_{2j+2n+2m-1}-J_{2j+2n+2m+1}\right) - J_{2j+2n+2m}\left(J_{2j+2m-1}-J_{2j+2m+1}\right)\right] + \delta_{n=0}.\label{eq:aat}
\end{eqnarray}
\begin{eqnarray}
\left\langle B_{j}(t)B_{j+n}(t)\right\rangle & = & - i\left(J_{2j+1}J_{2j+2n}-J_{2j}J_{2j+2n+1}\right) +i\sum_{m=1}^{2j+n}\left[J_{2j-2m+1}J_{2j+2n-2m} - J_{2j-2m}J_{2j+2n-2m+1}\right]\nonumber\\
& -& i\sum_{m>0}\left[J_{2j+2m+2n}\left(J_{2j+2m+1}-J_{2j+2m-1}\right) - J_{2j+2m}\left(J_{2j+2n+2m+1}-J_{2j+2m+2n-1}\right)\right]  -\delta_{n=0}\label{eq:bbt}
\end{eqnarray}
\end{widetext}
In the above expressions we have suppressed the arguments $2Jt\,$ common to every Bessel function. Our main interest is in the behavior of local observables at long enough times that a central region of essentially flat magnetization has formed, corresponding to the limit $|j|,|n| \ll Jt\,$. As shown in Appendix~\ref{sec:bessel}, Eqs.~(\ref{eq:bat})-(\ref{eq:bbt}) reduce in this limit to
\begin{eqnarray}
\lim_{t\rightarrow \infty}\left\langle B_{j}(t)A_{j+n}(t)\right\rangle & = & 0,\label{eq:baness}\\
\lim_{t\rightarrow \infty}\left\langle A_{j}(t)A_{j+n}(t)\right\rangle & = & \delta_{n=0} + \frac{i}{\pi}\frac{4n(-1)^{n+1}}{4n^{2} -1}\label{eq:aaness},\\
\lim_{t\rightarrow \infty}\left\langle B_{j}(t)B_{j+n}(t)\right\rangle & = & -\lim_{t\rightarrow \infty}\left\langle A_{j}(t)A_{j+n}(t)\right\rangle\label{eq:bbness}.
\end{eqnarray}
Strictly speaking, Eqs.~(\ref{eq:bat}) and (\ref{eq:bbt}) are valid for $2j+n\geq 0$. For $2j+n<0\,$ corresponding expressions may be derived which also reduce to Eqs.~(\ref{eq:baness})-(\ref{eq:bbness}) in the limit $|j|,|n| \ll Jt$. For the non-interacting problem under consideration, Wick's theorem may then be used to reduce any local observable to expressions involving Eqs.~(\ref{eq:baness})-(\ref{eq:bbness}).
\subsection{Observables}
For the critical transverse-Ising model, we may obtain explicit expressions for the current, magnetization and $\mathcal{C}_{n}^{\nu\nu}(j,t),$ correlation functions in the long-time limit. Furthermore, due to the simple structure of Eqs.~(\ref{eq:baness})-(\ref{eq:bbness}), we may reduce the Pfaffian expression for $\mathcal{C}_{n}^{xx}(j,t)\,$ to a determinant of a particular skew-symmetric $n\times n\,$ matrix, which is no more difficult to evaluate than the equilibrium correlation function. To this end, the magnetization in the center region $ |j| \ll Jt\,$ is seen to relax to zero,
\begin{equation}
\lim_{t\rightarrow \infty}\left\langle \hat{S}_{j}(t)\right\rangle = \frac{1}{2}\lim_{t\rightarrow\infty}\left\langle B_{j}(t)A_{j}(t)\right\rangle = 0.
\end{equation}
As we shall demonstrate in the next section, the steady-state magnetization does {\it not} vanish in the central region for nonzero $h\,$ and arbitrary choices of domain-wall height $m_{0}<\frac{1}{2}$. Interestingly, the magnetization far from the central region relaxes to half its initial value,
\begin{equation}
\left\langle \hat{S}_{j}(t)\right\rangle_{|j|  \gg Jt \gg 0} = -\frac{1}{4}\mbox{sgn}(j),
\end{equation}
a result which has been previously obtained \cite{Karevski2002}. Since total $c$-fermion number (the magnetization) is no longer conserved for $\gamma \neq 0$, this suggests that the $c$-fermion number operators share some nonzero overlap with the $\eta$-fermion number operator which is conserved under the final Hamiltonian. After the initial magnetization relaxation, the domain wall dynamics appears similar in spirit to the isotropic model, and one might reasonably expect a nonzero current in the center. Indeed, one finds
\begin{eqnarray}
\lim_{t\rightarrow \infty}\left\langle \hat{\mathcal{J}}^{z}_{j}(t)\right\rangle & = & \frac{J}{2}\lim_{t\rightarrow \infty}\mbox{Im}\left\langle A_{j}(t)A_{j+1}(t)\right\rangle,\\
& = & \frac{2J}{3\pi}\label{eq:ticur}.
\end{eqnarray}
Equation~(\ref{eq:ticur}) is interesting for two reasons. First, despite the domain wall height decreasing to half its initial value after the initial relaxation, the current is only reduced to $\frac{2}{3}\,$ of its value in the isotropic $XX$ spin chain, $\frac{J}{\pi}$. Additionally, if one instead considers a current-carrying, ``boosted-Fermi-sea,'' initial state \cite{Antal1998,Sabetta2013} with maximum current allowed by the lattice, it follows that the initial current remains conserved in the long-time limit for any values of $h,\gamma$. That is for
\begin{equation}
\left\langle c_{k}^{\dagger}c_{k}\right\rangle = \left\{\begin{array}{cc} 1 & k\in \left(-\frac{\pi}{2}+\phi, \frac{\pi}{2} + \phi\right), \\ 0 & \mbox{otherwise}\end{array}\right.,
\end{equation}
with $\phi = \frac{\pi}{2}$, Eq.~(\ref{eq:cur1}) gives an initial current $\left\langle \mathcal{J}^{z}_{j}\right\rangle = \frac{J}{\pi}$. Performing time evolution with this initial state and using Eqs.~(\ref{eq:xyt})-(\ref{eq:gt}) one finds $\left\langle \hat{\mathcal{J}}^{z}_{j}(t)\right\rangle \rightarrow \frac{J}{\pi}$. One can imagine carrying out the quench in a two-step procedure, first relaxing the spatially varying magnetic field and leaving $h = 0$, $\gamma = 0$. The domain wall spreads ballistically, giving rise to a central steady-state current $\mathcal{J}_{0} = \frac{J}{\pi}$. After a long time has passed, a sudden quench in anisotropy and magnetic field $h \rightarrow J$, $\gamma \rightarrow 1\,$ will result in some transient dynamics, but the current at long times will average to $\mathcal{J}_{0}$. Alternatively, if the spatially varying magnetic field is switched off {\it at the same time} the anisotropy and homogeneous field are turned on, the current will approach $\frac{2}{3}\mathcal{J}_{0}\,$ at long times. In this way, the precise details of the quench protocol affect the long-time steady state even in the absence of interactions.

\begin{figure}
\begin{center}
\includegraphics[totalheight=6.25cm]{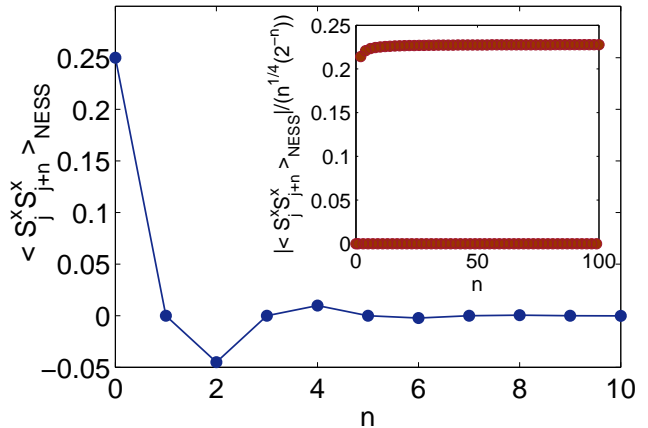}
\caption{(Color online) Long-time limit of transverse correlations $\mathcal{C}^{xx}_{j,j+n}(t)\,$ for $Jt \gg |j|,|n|\,$ after a quench to the transverse-Ising limit with an initial magnetization domain wall of height $m_{0} = \frac{1}{2}$. Oscillations are superimposed on an overall exponential decay (inset). The system retains memory of its initial state through both the short wave-length of oscillations and the equivalence between $\mathcal{C}^{xx}_{j,j+n}(t)\,$ and  $\mathcal{C}^{yy}_{j,j+n}(t)\,$ (not shown).}
\label{fig:cxx}
\end{center}
\end{figure}

The long-time limit of the correlation functions in the central region may also be investigated. Using Eq.~(\ref{eq:czz}) and Eqs.~(\ref{eq:baness})-(\ref{eq:bbness}) we obtain a power-law decay for $\mathcal{C}_{n}^{zz}$,
\begin{equation}
\lim_{t\rightarrow\infty}\mathcal{C}_{n}^{zz}(j,t) = -\frac{4n^{2}}{\pi^{2}(4n^{2}-1)^{2}}\;\;(n>0).
\end{equation}
Turning attention to the transverse correlations, Eq.~(\ref{eq:cxxpfaf}) for $\mathcal{C}_{n}^{xx}(j,t)\,$ may be organized as \cite{Barouch2}
\begin{equation}
\left[\mathcal{C}_{n}^{xx}(j,t)\right]^{2} = \frac{1}{16}\mbox{det}\left(\begin{array}{cc} {\bf S} & {\bf G}\\ {\bf -G}^{T} & {\bf Q}\end{array}\right),\label{eq:cxxness}
\end{equation}
where 
\begin{eqnarray}
G_{lk} & = & \left\langle B_{l+j-1}(t)A_{k+j}(t)\right\rangle,\\
S_{lk} & = & \left\langle B_{l+j-1}(t)B_{k+j-1}(t)\right\rangle\;\; (k>l),\\
Q_{lk} & = & \left\langle A_{l+j}(t)A_{k+j}(t)\right\rangle \;\;(k>l).
\end{eqnarray}
Here $S_{kl} = -S_{lk}\,$ and $Q_{kl} = -Q_{lk}$. Furthermore, according to Eq.~(\ref{eq:baness}) $G_{lk} = 0\,$ in the central region in the long-time limit. Equations~(\ref{eq:aaness}) and (\ref{eq:bbness}) indicate that $S_{lk}\,$ and $Q_{lk}\,$ depend on $l\,$ and $k\,$ only through the combination $l-k$. Using the skew-symmetric nature of {\bf Q}, Eq.~(\ref{eq:cxxness}) may be reduced to 
\begin{eqnarray}
\lim_{t\rightarrow\infty}\mathcal{C}_{n}^{xx}(j,t) & = & \frac{i^{n}}{4}\mbox{det}{\bf Q},\\
& = & \frac{i^{n}}{4}\left|\begin{array}{cccc} q_{0} & q_{1} & \cdots & q_{n-1} \\ q_{-1} & q_{0} & \cdots & q_{n-2}\\ \vdots & \vdots & \ddots & \vdots \\ q_{-n+1} & q_{-n+2} & \cdots & q_{0}\end{array}\right|,\label{eq:cdet}
\end{eqnarray}
where $q_{n} = \frac{4n(-1)^{n-1}}{\pi(4n^{2}-1)}$. The matrix in Eq.~(\ref{eq:cdet}) is an example of a Toeplitz matrix, satisfying $Q_{lk} = q_{k-l}$. Due to the antisymmetry of {\bf Q} through $q_{n} = -q_{-n}$, only even values of $n\,$ result in a non-vanishing $\mathcal{C}_{n}^{xx}(j,t)$. The phase $i^{n}\,$ may be factored out and is purely real, with $i^{n} \rightarrow \cos\left(\frac{\pi n}{2}\right)\,$ for even $n$. Our interest is in extracting the asymptotic behavior of this determinant for large $n$, which is facilitated by defining a generating function, $\tilde{q}(k)$, satisfying
\begin{equation}
\tilde{q}(k) = \sum_{n}e^{ikn}q_{n},\;\;\; q_{n} = \int_{-\pi}^{\pi}\frac{dk}{2\pi}e^{-ikn}\tilde{q}(k).
\end{equation}
According to the Fisher-Hartwig conjecture \cite{FH1,FH2,FH4}, if $\tilde{q}(k)\,$ has the form 
\begin{equation}
\tilde{q}(k) = f(k)\prod_{j=1}^{m}e^{i\beta_{j}(k-k_{j}-\pi\mbox{\scriptsize sgn}(k-k_{j}))}\left|2-2\cos(k-k_{j})\right|^{\alpha_{j}},
\end{equation}
with jump discontinuities occurring at isolated points $k = k_{j}\,$ characterized by exponents $\beta_{j}\,$ and zeros characterized by exponents $\alpha_{j}$, then the asymptotic form of the determinant determinant for large $n\,$ is given by
\begin{equation}
D_{n}\left[q\right] \sim \mathcal{M}e^{f_{0}n}\exp\left[\sum_{j=1}^{m}\left(\alpha_{j}^{2}-\beta_{j}^{2}\right)\ln n\right]\;\;\;(\mbox{as }n\rightarrow \infty),\label{eq:fhc}
\end{equation}
where $\mathcal{M}\,$ is a numerical constant independent of $n\,$ and
\begin{equation}
f_{0} = \int_{0}^{2\pi}\ln f(k) \frac{dk}{2\pi}.\label{eq:co}
\end{equation}
A Fourier transform of Eq.~(\ref{eq:aaness}) yields
\begin{eqnarray}
\tilde{q}(k) & = & -i\sin\left[\frac{1}{2}\left(2m\pi - k\right)\right],\\
& & (2m\pi < k <(2m+2)\pi),
\end{eqnarray}
for $m = 0,1,2,\cdots$. Taking $m = 0\,$ and accounting for the jump discontinuity at $k = \pi$, we may represent
\begin{equation}
\tilde{q}(k)  =  \frac{1}{2}t_{-\frac{1}{2}}(k)t_{-\frac{1}{2}}(k-\pi)|2-2\cos k|^{\frac{1}{2}},\label{eq:fhf}
\end{equation}
with $t_{\beta}(k) \equiv e^{i\beta(k-\pi\mbox{\scriptsize sgn}(k))}$. This representation makes explicit the two Fisher-Hartwig singularities arising at $k_{1}=0\,$ ($\alpha_{1} = \frac{1}{2}$, $\beta_{1} = -\frac{1}{2}$) and $k_{2}  = \pi\,$ ($\alpha_{2} = 0$, $\beta_{2}  =-\frac{1}{2}$). The numerical prefactor leads to exponential decay and,
\begin{equation}
f_{0} = \int_{0}^{2\pi}\ln \frac{1}{2}\frac{dk}{2\pi} = -\ln2.
\end{equation}
Employing the straightforward methods in Ref.~\onlinecite{Ovchinnikov2007}, we may calculate the constant
\begin{equation}
\mathcal{M} = \frac{e^{\frac{1}{4}}2^{\frac{7}{12}}}{\mathcal{A}^{3}}\approx 0.912,
\end{equation}
where $\mathcal{A} \approx 1.2824\,$ is the Glaisher-Kinkelin constant. Collecting all the pieces using Eq.~(\ref{eq:fhc}) yields
\begin{equation}
\lim_{t\rightarrow\infty}\mathcal{C}_{n}^{xx}(j,t) \sim \frac{e^{\frac{1}{4}}2^{\frac{7}{12}}}{4\mathcal{A}^{3}}\cos\left(\frac{\pi n}{2}\right)n^{-\frac{1}{4}}2^{-n}\;\;(\mbox{as }n\rightarrow \infty).
\end{equation}
The $\cos\left(\frac{\pi n}{2}\right)\,$ results from $i^{n}\,$ under the condition that $n\,$ is even, and $\frac{1}{4}\mathcal{M} \approx 0.228$. Previously a similar result was obtained \cite{Sengupta2004} by quenching the transverse-Ising model from the ground state of $h = \infty\,$ to the critical point $h = J$. In that case, it was found that $\mathcal{C}^{xx}_{n}(t) \rightarrow \left(\frac{1}{2}\right)^{n}\,$ exactly. Here we find a power-law decay factor in addition to the exponential decay. Curiously, the same exponent appears in the ground state correlation function, which is given by purely power-law decay [Eq.~(\ref{eq:xtieq})]. A graphical depiction of $\mathcal{C}^{xx}_{n}(j,t)\,$ at long times is shown in Fig.~\ref{fig:cxx}. The Pfaffian expression for $\mathcal{C}^{yy}_{n}(j,t)\,$ may be evaluated numerically or reorganized into the basic structure of Eq.~(\ref{eq:cxxness}). From either approach, it follows that
\begin{equation}
\lim_{t\rightarrow\infty }\mathcal{C}^{xx}_{n}(j,t) = \lim_{t\rightarrow\infty }\mathcal{C}^{yy}_{n}(j,t),
\end{equation}
despite time evolution being generated by the anisotropic $XY$ model at the Ising critical point. This equality in transverse correlations is reflective of the rotational symmetry present in the initial state.
\subsection{Homogeneous quench}
As a specific point of comparison, we may also calculate $\mathcal{C}^{xx}_{n}(t)\,$ in the long-time limit after a rapid anisotropy quench from the ground state of the $XX$ model to the same critical transverse-Ising Hamiltonian. The motivation behind this brief digression is to understand the effects of the anisotropy quench separately from the effects due to time evolution of a spatially inhomogeneous initial state. That is, in this section we briefly consider the simpler quench in which the system begins in the ground state of
\begin{equation}
\hat{H}_{0} = -J\sum_{j}\left[\hat{S}_{j}^{x}\hat{S}_{j+1}^{x} + \hat{S}_{j}^{y}\hat{S}_{j+1}^{y}\right].
\end{equation}
This ground state is given by
\begin{equation}
\left|\Psi_{0}\right\rangle = \prod_{|k|<\frac{\pi}{2}}c_{k}^{\dagger}\left|0\right\rangle,
\end{equation}
and time evolution takes place with respect to the critical transverse-Ising model, Eq.~(\ref{eq:xy}) with $\gamma = \frac{h}{J}=1$, as in the previous section. This situation is described by the same time-dependent operators, which may be written in terms of the Majorana operators,
\begin{eqnarray}
B_{j}(t) & = & \frac{1}{\sqrt{N}} \sum_{k}e^{ikj}\left[(f_{k}^{*}-g_{k})c_{-k}^{\dagger} - (f_{k}+g_{k})c_{k}\right],\\
A_{j}(t) & = & \frac{1}{\sqrt{N}}\sum_{k}e^{ikj}\left[(f_{k}^{*}+g_{k})c_{-k}^{\dagger} + (f_{k}-g_{k})c_{k}\right],
\end{eqnarray}
with the initial occupation numbers
\begin{equation}
\left\langle c_{k}^{\dagger}c_{k'}\right\rangle = \delta_{kk'}\left[\frac{1}{2}+\mbox{sgn}\left(\frac{\pi}{2}+k\right)\mbox{sgn}\left(\frac{\pi}{2}-k\right)\right].
\end{equation}
We have suppressed the time-dependence in $f_{k}(t)$, $g_{k}(t)$ for brevity. Taking the long-time limit, one finds
\begin{eqnarray}
\lim_{t\rightarrow\infty}\left\langle A_{j}A_{j+n}\right\rangle & = & - \lim_{t\rightarrow\infty}\left\langle B_{j}B_{j+n}\right\rangle = \delta_{n=0},\\
\lim_{t\rightarrow\infty}\left\langle B_{j}A_{j+n}\right\rangle & = & \int_{-\pi}^{\pi}\frac{dk}{2\pi}e^{-ikn}\left[\frac{1}{2}+\frac{1}{2}e^{2i\theta_{k}}\right]\nonumber\\
& \times & \left(2\left\langle c_{k}^{\dagger}c_{k}\right\rangle - 1\right),
\end{eqnarray}
Despite being far from equilibrium, the structure of Eq.~(\ref{eq:cxxness}) reduces to a Toeplitz determinant as in equilibrium,
\begin{eqnarray}
\lim_{t\rightarrow\infty}\mathcal{C}^{xx}_{n}(t) & = & \frac{1}{4}\left|\begin{array}{cccc} g_{1} & g_{0} & \cdots & g_{-n+2}\\ g_{2} & g_{1} & \cdots & g_{-n+1} \\ \vdots & \vdots & \ddots & \vdots\\ g_{n} & g_{-n+1} & \cdots & g_{1}\end{array}\right|,
\end{eqnarray}
with
\begin{eqnarray}
g_{n} & = & \int_{-\pi}^{\pi}\frac{dk}{2\pi}e^{-ikn}\tilde{g}(k),\\
\tilde{g}(k) & = & \frac{e^{-ik}}{2}\left(1+e^{2i\theta_{k}}\right)\mbox{sgn}\left(\frac{\pi}{2}+k\right)\mbox{sgn}\left(\frac{\pi}{2}-k\right).
\end{eqnarray}
First we note that
\begin{eqnarray}
e^{-ik}& \mbox{sgn} & \left(\frac{\pi}{2}+k\right)\mbox{sgn}\left(\frac{\pi}{2}-k\right)\nonumber\\
& = & t_{-\frac{1}{2}}\left(k-\frac{\pi}{2}\right)t_{-\frac{1}{2}}\left(k+\frac{\pi}{2}\right).
\end{eqnarray}
So far the analysis is valid for arbitrary $0 \leq \gamma, \frac{h}{J} \leq 1$. Specializing to $\gamma = \frac{h}{J} = 1$, one finds the remaining factor can be written
\begin{eqnarray}
\left(1-e^{ik}\right) & = & -i e^{\frac{ik}{2}}\mbox{sgn}(k)|2-2\cos k|^{\frac{1}{2}}\\
& = & t_{\frac{1}{2}}\left(k\right)|2-2\cos k|^{\frac{1}{2}},
\end{eqnarray}
so that we have a generating function with three Fisher-Hartwig singularities
\begin{eqnarray}
\tilde{g}(k) & = & \frac{1}{2}t_{\frac{1}{2}}(k)t_{-\frac{1}{2}}\left(k-\frac{\pi}{2}\right)t_{-\frac{1}{2}}\left(k+\frac{\pi}{2}\right)\left|2-2\cos k\right|^{\frac{1}{2}}.
\end{eqnarray}
The appearance of an additional jump factor $t_{\beta}\,$ here compared to Eq.~(\ref{eq:fhf}) results in an enhancement of the power-law decay factor when rapidly changing $h \rightarrow J\,$ and increasing the anisotropy from zero to unity. As in the previous Subsection, one may calculate the constant $\mathcal{M}'\,$ by standard procedures, giving
\begin{eqnarray}
\lim_{t\rightarrow\infty}\mathcal{C}^{xx}_{n}(t) & \sim  & \frac{1}{4}\mathcal{M}'n^{-\frac{1}{2}}2^{-n}\\
& = & \frac{e^{\frac{1}{2}}2^{-\frac{5}{6}}}{\mathcal{A}^{6}}n^{-\frac{1}{2}}2^{-n}\;\;\;(\mbox{as }n\rightarrow \infty),\label{eq:cxxhomquench}
\end{eqnarray}
with $\frac{1}{4}\mathcal{M}' \approx 0.2075$. 
\begin{figure}
\begin{center}
\includegraphics[totalheight=6.25cm]{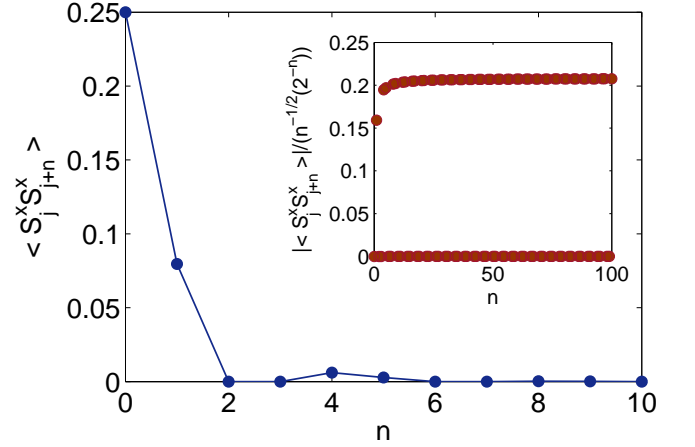}
\caption{(Color online) Long-time limit of transverse correlations $\mathcal{C}^{xx}_{j,j+n}(t)\,$ for $Jt \gg |j|,|n|\,$ for the ground state of the $XX$ model after a rapid anisotropy and magnetic field quench to the critical transverse-Ising limit $\gamma = \frac{h}{J} = 1$. Note that $\mathcal{C}^{xx}_{n} = 0\,$ for $\mbox{mod}(n,4) = 2,3$, which is not captured by Eq.~(\ref{eq:cxxhomquench}). Such oscillations are not present in $\mathcal{C}^{yy}$, which possesses the same decay envelope in Eq.~(\ref{eq:cxxhomquench}).}
\label{fig:cxx2}
\end{center}
\end{figure}
We note that while Eq.~(\ref{eq:cxxhomquench}) accurately captures the asymptotic decay envelope in $\mathcal{C}^{xx}_{n}(t)\,$ at long times, some disclaimers regarding the validity of this prediction should be mentioned. As can be seen from Fig.~\ref{fig:cxx2}, nontrivial oscillatory behavior is present in the nonequilibrium steady-state correlations. Specifically, $\mathcal{C}^{xx}_{n}(t)\rightarrow 0\,$ for $\mbox{mod}(n,4)=2,3$. This behavior is not at all captured at the level of analysis pursued here. Our main interest lies in extracting the asymptotic decay envelope [Eq.~(\ref{eq:cxxhomquench})], which agrees well with the numerical computation of the corresponding determinant for numerically accessible $n$. Interestingly, these oscillations are not present in $\mathcal{C}^{yy}_{n}$, which also decays according to Eq.~(\ref{eq:cxxhomquench}). It is only because the factor of $i\,$ may be factored out from each entry in the determinant in Eq.~(\ref{eq:cdet}) that the non-equilibrium oscillations arising from the initial domain wall are captured in our approach. A rigorous treatment regarding the applicability and possible extensions to the present form of the Fisher-Hartwig conjecture lies beyond the scope of this work. However, it appears clear from the present treatment that the power-law factor present in correlations arises from discontinuities in the momentum distribution at zero temperature. Notably, discontinuities are absent in the fully polarized initial state considered in a similar quench in Ref.~\onlinecite{Sengupta2004}. In that study, correlation functions were computed exactly for the case $\frac{h}{J}=1$, showing exponential decay without additional power-law factors. Additionally, we note a recent and intriguing work \cite{Vidmar2015} in which it was argued that power-law decay far from equilibrium can be understood in certain cases by viewing the time-dependent many-body state as the ground state of an emergent Hamiltonian, in which time appears as a parameter. The dephasing effects due to the anisotropy quench considered here appear too severe for the system to retain purely power-law correlations with respect to the two-point correlation functions.

\section{Nonequilibrium steady state as homogeneous subsystem}
\label{sec:ness}
Though Eqs.~(\ref{eq:Fmt}) and (\ref{eq:Gmt}) constitute a formal solution to the problem, the behavior of interest takes place within a regime in which $t\gg |j|,|n|\,$ (for local operators involving sites between $j\,$ and $j+n$), so that a homogeneous central region forms which is coupled to the remainder of the system through a spin current moving magnetization from the left edge to the right edge. In this section, we obtain an expression for the nonequilibrium steady state (NESS) occupation numbers in this central region by extending the analysis presented in Ref.~\onlinecite{Sabetta2013} to include anisotropy. The fundamental object of interest is the time-dependent occupation matrix,
\begin{eqnarray}
& & \left\langle c_{m}^{\dagger}(t)c_{n}(t)\right\rangle = \int \frac{dk}{2\pi}e^{-ik(m-n)} \int \frac{dq}{2\pi}e^{-i(m+n)q/2}\nonumber\\
& \times & \left\langle \left[f^{*}_{k_{+}}(t)c_{k_{+}}^{\dagger} + g_{k_{+}}(t)c_{-k_{+}}\right]\left[f_{k_{-}}(t)c_{k_{-}} + g_{k_{-}}(t)c_{-k_{-}}^{\dagger}\right]\right\rangle \\
& = & \int \frac{dk}{2\pi}e^{-ik(m-n)} G(k),
\end{eqnarray}
with
\begin{eqnarray}
G(k) & \equiv & \left\langle c_{k}^{\dagger}c_{k}\right\rangle_{\mbox{\scriptsize NESS}}\\
& = & \sin^{2}\theta_{k}\sin^{2}\left(\epsilon_{k}t\right) \nonumber\\
& + & \int \frac{dq}{2\pi}e^{i(m+n)q/2}\left[f^{*}_{k_{+}}(t)f_{k_{-}}(t)\left\langle c_{k_{+}}^{\dagger}c_{k_{-}}\right\rangle\right.\nonumber\\
& - & \left. g_{k_{+}}(t)g_{k_{-}}(t)\left\langle c_{-k_{-}}^{\dagger}c_{-k_{+}}\right\rangle\right]\label{eq:GpNess}
\end{eqnarray}
Here $k_{\pm} \equiv k \pm \frac{q}{2}$. The long-time limit of observables for the $XX$ model with an initial magnetization domain wall of reduced height $m_{0} < \frac{1}{2}\,$ has previously been explored \cite{Antal1999,Lancaster2010a,Sabetta2013}, and in this section we sketch some new features which arise in the presence of nonzero anisotropy and external magnetic field for $m_{0} = \frac{1}{2}\,$ and $m_{0}<\frac{1}{2}$. The domain-wall height in the ground state of Eq.~(\ref{eq:hinit}) is always $m_{0} = \frac{1}{2}$, as the width is the only tunable parameter. In the limit of a sharp jump in magnetization $\mathcal{\alpha}\rightarrow 0$, this state can be viewed as the union of two homogeneous subsystems, the left being filled up to Fermi momentum $k_{L} = k^{+} = \pi\,$ and the right being entirely empty, $k_{R} = k^{-} = 0$. In Ref.~\onlinecite{Antal1999}, it was observed that this picture can be slightly modified to create domain walls of arbitrary $m_{0}<\frac{1}{2}\,$ by letting
\begin{equation}
k^{\pm} = \frac{\pi}{2} \pm \pi m_{0}.
\end{equation}
With nonzero anisotropy, we can expand Eq.~(\ref{eq:GpNess}) for small $q$, obtaining
\begin{eqnarray}
f_{k_{+}}(t)f_{k_{-}}(t) & \simeq & \frac{1}{2}\cos\left(\delta\epsilon t\right)\left(1+  \cos^{2}\theta_{k}\right) \nonumber\\ 
& + & i\cos\theta_{k}\sin\left(\delta\epsilon t\right),\\
g_{k_{+}}(t)g_{k_{-}}(t) & \simeq & \frac{1}{2}\cos\left(\delta\epsilon t\right) \sin^{2}\theta_{k},
\end{eqnarray}
where $\delta\epsilon t \equiv \left(\epsilon_{k+q/2}-\epsilon_{k-q/2}\right) t \simeq v_{k}qt$, with $v_{k} = \partial_{k}\epsilon_{k}\,$ the mode velocity. The remainder of the calculation closely follows the procedure outlined in Ref.~\onlinecite{Sabetta2013}, and we only highlight the crucial steps in what follows. Since the above integral is saturated by values of $\delta\epsilon \sim \mathcal{O}(t^{-1})$, we formally change variables to
\begin{equation}
u = \delta\epsilon t = v_{k}qt,
\end{equation}
giving
\begin{eqnarray}
G(k) & = & \frac{1}{2}\sin^{2}\theta_{k} \nonumber\\
& + & \frac{1}{2}\left(1+\cos^{2}\theta_{k}\right)\int_{-\infty}^{\infty}\frac{du}{2\pi|v_{k}|t}\cos(u)\left\langle c_{k_{+}}^{\dagger}c_{k_{-}}\right\rangle \nonumber\\
& + & i\cos\theta_{k}\int_{-\infty}^{\infty}\frac{du}{2\pi|v_{k}|t}\sin(u)  \left\langle c_{k_{+}}^{\dagger}c_{k_{-}}\right\rangle \nonumber\\
& - & \frac{1}{2}\sin^{2}\theta_{k}\int_{-\infty}^{\infty}\frac{du}{2\pi |v_{k}|t}\cos(u)\left\langle c_{-k_{-}}^{\dagger}c_{-k_{+}}\right\rangle.
\end{eqnarray}
In the presence of anisotropy (nonzero $\theta_{k}$) there is also an anomalous average,
\begin{eqnarray}
F(k) & = & \left\langle c_{k}c_{-k}\right\rangle_{\mbox{\scriptsize NESS}}\\
& = &  -\frac{i}{2}\sin(2\theta_{k}) \int_{-\infty}^{\infty}\frac{du}{2\pi|v_{k}|t}\nonumber\\
& \times & e^{-i(m+n)q/2}\frac{1}{2}\left[-\left\langle c_{k_{+}}c_{k_{-}}^{\dagger}\right\rangle + \left\langle c_{-k_{+}}^{\dagger}c_{-k_{-}}\right\rangle\right].
\end{eqnarray}
These expressions may be evaluated by using the ansatz \cite{Sabetta2013} for the initial off-diagonal momentum correlations,
\begin{equation}
\left\langle c_{k+\frac{q}{2}}^{\dagger}c_{k-\frac{q}{2}}\right\rangle \approx \Theta(k^{-}-|k|)\frac{i}{q+i0^{+}} -\Theta(k^{+}-|k|)\frac{i}{q+i0^{-}}.
\end{equation}
This procedure is straightforward and yields the following expressions
\begin{eqnarray}
G(k) & = & \frac{1}{2}\sin^{2}\theta_{k} \nonumber\\
& + & \frac{1}{4}\left(1+\cos(2\theta_{k})\right)\left[\Theta(k^{+}-|k|) + \Theta(k^{-}-|k|)\right]\nonumber\\
& + & \frac{1}{2}\Theta(k)\cos\theta_{k}\Theta(k-k^{-})\Theta(k^{+}-k) \nonumber\\ 
& - &  \frac{1}{2}\Theta(-k)\cos\theta_{k}\Theta(-k^{-}-k)\Theta(k^{+}+k),\label{eq:Gness}\\
F(p) & = & \frac{i}{4}\sin(2\theta_{k}) \nonumber\\
& - & \frac{i}{4}\sin(2\theta_{k})\left[\Theta(k^{+}-|k|) + \Theta(k^{-}-|k|)\right].\label{eq:Fness}
\end{eqnarray}
For $k^{+}=k^{-} = k_{F}$, these expressions reduce to those corresponding to a homogeneous quench from the $XX$ model ground state to the $XY$ Hamiltonian with Bogoliubov angle $\theta_{p}$ predicted from the generalized Gibbs ensemble \cite{Cazalilla2006,Iucci2009,Cazalilla2012},
\begin{eqnarray}
G_{0}(k) & \xrightarrow{m_{0}\rightarrow0} & \frac{1}{2}\sin^{2}\theta_{k} + \frac{1}{2}\left(1+\cos^{2}\theta_{k}\right)n_{k},\\
F_{0}(k) & \xrightarrow{m_{0}\rightarrow0} & \frac{i}{4}\sin(2\theta_{k}) - \frac{i}{2}\sin(2\theta_{k})n_{k},
\end{eqnarray}
where $n_{k} = \Theta\left(k_{F}-|k|\right)\,$ are the occupation numbers for the ground state of the $XX$ model with zero field. From Eqs.~(\ref{eq:Gness}) and (\ref{eq:Fness}) we proceed to directly compute observables within the emergent non-equilibrium steady state with little more effort than required for computing the corresponding ground state observables.
\subsection{Observables for $m_{0} = \frac{1}{2}$}
\begin{figure}
\begin{center}
\includegraphics[totalheight=6.25cm]{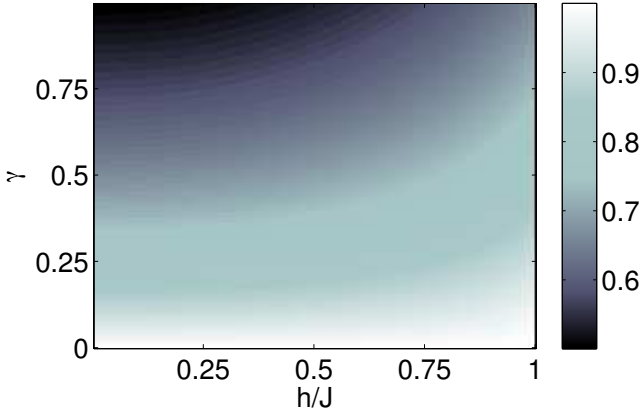}
\caption{(Color online) Residual current as a fraction of that which appears in the $XX\,$ model at long times with the full domain-wall initial state, $m_{0} = \frac{1}{2}$. Here $\mathcal{J}_{0} = \frac{J}{\pi}\,$ is the corresponding current for $\gamma = \frac{h}{J} = 0$.}
\label{fig:nesscur}
\end{center}
\end{figure}

Equations~(\ref{eq:Gness}) and (\ref{eq:Fness}) give the required information for the basic contractions within the nonequilibrium steady state which forms at long times with $|j|, |n| \ll Jt$. Explicitly,
\begin{eqnarray}
\left\langle c_{j}^{\dagger}c_{j+n}\right\rangle_{\mbox{\scriptsize NESS}} & = & \int_{-\pi}^{\pi}\frac{dk}{2\pi}e^{ikn}G(k),\label{eq:cjdness}\\
\left\langle c_{j}c_{j+n}\right\rangle_{\mbox{\scriptsize NESS}} & = & \int_{-\pi}^{\pi}\frac{dk}{2\pi}e^{-ikn}F(k)\label{eq:cjness}.
\end{eqnarray}
For the maximum jump in magnetization across the initial domain-wall profile ($m_{0} = \frac{1}{2}$, $k^{+} = \pi$, $k^{-} = 0$) Eqs.~(\ref{eq:cjdness}) and (\ref{eq:cjness}) lead to 
\begin{eqnarray}
\left\langle B_{j}A_{j+n}\right\rangle_{\mbox{\scriptsize NESS}} & = & 0,\label{eq:bass}\\
\left\langle A_{j}A_{j+n}\right\rangle_{\mbox{\scriptsize NESS}} &  = &  \int_{-\pi}^{\pi}\frac{dk}{2\pi}e^{-ikn}\mbox{sgn}(k)\cos\theta_{k}\nonumber\\
&  & + \delta_{n=0},\label{eq:aass}\\
\left\langle B_{j}B_{j+n}\right\rangle_{\mbox{\scriptsize NESS}} & = &-\left\langle A_{j}A_{j+n}\right\rangle_{\mbox{\scriptsize NESS}}.\label{eq:bbss}
\end{eqnarray}
Eqs.~(\ref{eq:bass})-(\ref{eq:bbss}) hold in the long-time limit for any choice of $\gamma\,$ and $h$, which means that the transverse correlation function may be calculated in this steady state in terms of a standard Toeplitz determinant of the form in Eq.~(\ref{eq:cdet}) with
\begin{eqnarray}
q_{n} & = & i\int_{0}^{\pi}\frac{dk}{\pi} \frac{\sin (kn) \left|\cos k - \frac{h}{J}\right|}{\sqrt{\left(\cos k - \frac{h}{J}\right)^{2} + \gamma^{2}\sin^{2}k}}\label{eq:cn}\\
& \equiv & \int_{-\pi}^{\pi}\frac{dk}{2\pi }e^{-ikn}\tilde{c}(k),\\
\tilde{q}(k) & = & \frac{\mbox{sgn}(k)\left|\cos k - \frac{h}{J}\right|}{\sqrt{\left(\cos k - \frac{h}{J}\right)^{2} + \gamma^{2}\sin^{2}k}}.
\end{eqnarray}
As with the $XX\,$ model and the transverse-Ising limit, the antisymmetry of Eq.~(\ref{eq:cn}) with respect to $n\,$ forces $\mathcal{C}^{xx}_{\mbox{\scriptsize NESS}}(n) = 0\,$ for odd $n$. Combining this with the imaginary prefactor, we have a resulting phase factor
\begin{equation}
\mathcal{C}^{xx}_{\mbox{\scriptsize NESS}}(n) \propto \cos\left(\frac{\pi n}{2}\right),\label{eq:cxxcos2}
\end{equation}
for all choices of $\gamma\,$ and $h$. In the context of the $XX\,$ model, these oscillations may be viewed as arising due to a conserved spin current in the nonequilibrium steady state. Here we find exactly the same wavelength of oscillations and a reduced current (see below) which depends on the details of the anisotropy and the magnetic field through $\cos\theta_{k}$. That is, unlike the special isotropic case, the oscillations in $\mathcal{C}^{xx}_{n}\,$ are not determined by the long-time limit of $\left\langle \hat{\mathcal{J}}^{z}_{j}(t)\right\rangle$, which depends strongly on $\gamma\,$ and $h$. Figure~\ref{fig:nesscur} depicts the behavior of the long-time current for $0 \leq \gamma, \frac{h}{J} < 1$. Note that in all cases Eq.~(\ref{eq:cxxcos2}) implies the same wavelength for spatial oscillations in the two-point function. 

\begin{figure}
\begin{center}
\includegraphics[totalheight=6.25cm]{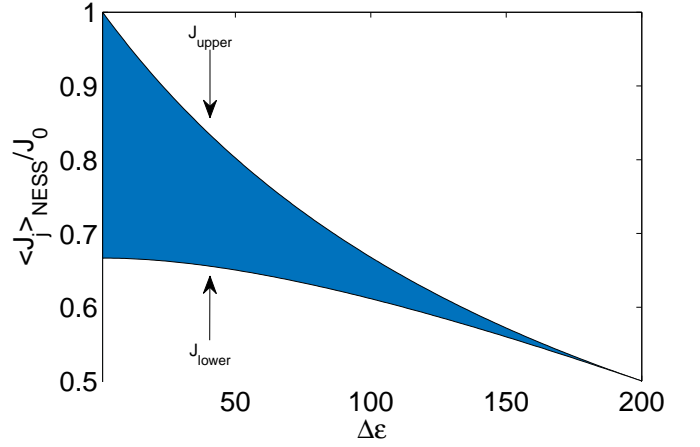}
\caption{(Color online) Dependence of ratio of residual current to $\mathcal{J}_{0}\,$ on energy gap $\Delta = 2\gamma\sqrt{1-\frac{h^{2}}{J^{2}}}\,$ in the final Hamiltonian.}
\label{fig:nesscur2}
\end{center}
\end{figure}

The steady-state current may be calculated by substituting Eqs.~(\ref{eq:Gness}) and (\ref{eq:cjdness}) into Eq.~(\ref{eq:cur1}), yielding
\begin{equation}
\left\langle \hat{\mathcal{J}}^{z}_{j}\right\rangle_{\mbox{\scriptsize NESS}} = J\int_{k^{-}}^{k^{+}}\frac{dk}{2\pi}\sin k \cos\theta_{k}.\label{eq:nesscur}
\end{equation}
Setting $k^{-} = 0$, $k^{+} = \pi$ and using $\cos\theta_{k} = \left|\sin \frac{k}{2}\right|$, we recover Eq.~(\ref{eq:ticur}) as a special case for the largest possible domain-wall state evolving under the critical transverse-Ising Hamiltonian. Equation~(\ref{eq:nesscur}) may be integrated numerically for arbitrary choices of $h\,$ and $\gamma$, as depicted in Fig.~\ref{fig:nesscur}. For nonzero anisotropy, the spectrum in Eq.~(\ref{eq:xydisp}) possesses an energy gap at $k_{F} = \cos^{-1}\frac{h}{J}\,$ of magnitude
\begin{equation}
\Delta \epsilon = 2J\gamma\sqrt{1-\left(\frac{h}{J}\right)^{2}}.
\end{equation}
One might expect from general considerations that the persistent current shows some tendency to decrease for larger energy gaps $\Delta \equiv \frac{\Delta \epsilon}{J}$. Collecting the values in Fig.~\ref{fig:nesscur} and plotting them as a function of $\Delta$, one finds a continuous band of allowed current as a function of $\Delta\,$ which becomes narrower for increasing gap size $\Delta$, as depicted in Fig.~\ref{fig:nesscur2}. The points do not form a single line as $\gamma\,$ and $h\,$ may be independently varied, with different choices of $\left(\gamma, h\right)\,$ producing the same $\Delta$.
The upper and lower boundaries of the shaded region in Fig.~\ref{fig:nesscur2} correspond to $h = 0\,$ with $0 < \gamma <1\,$ and $\gamma = 1\,$ with $0 < \frac{h}{J} < 1$, respectively. Furthermore, the shape of the boundaries may be computed explicitly from Eq.~(\ref{eq:nesscur}) in the appropriate regimes giving
\begin{eqnarray}
\mathcal{J}_{\mbox{\scriptsize upper}}(\Delta) & = & \frac{ \mathcal{J}_{0}}{1+\frac{\Delta}{2}},\label{eq:jupper}\\
\mathcal{J}_{\mbox{\scriptsize lower}}(\Delta) & = & \frac{1}{3}\mathcal{J}_{0}\left[2 - \frac{\Delta^{2}}{4\left(1+\frac{\Delta}{2}\right)}\right],
\end{eqnarray}
where $\mathcal{J}_{0} = \frac{J}{\pi}\,$ is the current generated by the domain-wall initial state in the $XX\,$ model. For $h=0\,$ ($\mathcal{J}_{\mbox{\scriptsize upper}}$) the gap is given by $\Delta = 2\gamma$, whereas for $\gamma = 1\,$ ($\mathcal{J}_{\mbox{\scriptsize lower}}$) we have $\Delta = 2\sqrt{1-\frac{h^{2}}{J^{2}}}$. 

\subsection{Observables for $m_{0} < \frac{1}{2}$}
 \begin{figure}
\includegraphics[totalheight=6.25cm]{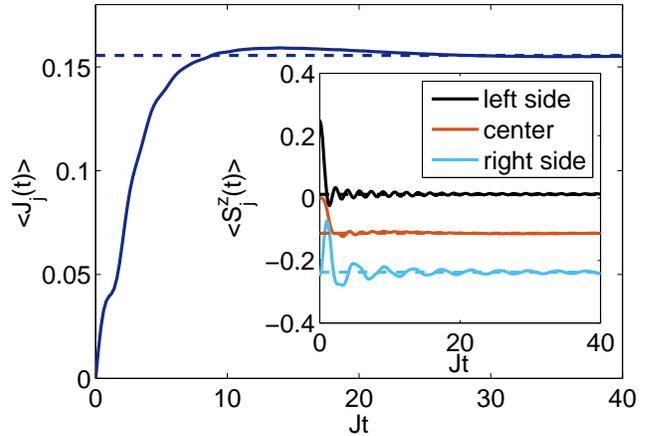}
\caption{(Color online) Current approaching nonequilibrium steady-state value for $m_{0} = 0.25$, $h = J$, $\gamma = 1\,$ (main). Magnetization also relaxes on either side of the domain wall and within the center region (inset). For $m_{0} < 0.5$, the central region relaxes to nonzero magnetization. Plots depict the average of instantaneous magnetization or current on six neighboring lattice sites obtained by applying Eqs.~(\ref{eq:Fmt}) and (\ref{eq:Gmt}) to the initial state studied in Ref.~\onlinecite{Antal1999}, which corresponds to a sharp domain wall of height $m_{0}<\frac{1}{2}$.}
\label{fig:nessapproach}
\end{figure}

We may employ the nonequilibrium steady state derived above to examine observables evaluated in the center region in the long-time limit analytically without having closed-form expressions for Eqs.(\ref{eq:Fmt}) and (\ref{eq:Gmt}) for arbitrary $\gamma\,$ and $\frac{h}{J}\,$ with $m_{0} < \frac{1}{2}$. The magnetization dynamics can be understood as a two-stage process in which the right and left sides of the system quickly relax according to a homogeneous quench. That is, far to the right or left of the origin, the system is only aware of the anisotropy quench and behaves accordingly. Each side may thus be described by a generalized Gibbs ensemble as follows. For $t>0\,$ the conserved quantities are
\begin{equation}
\mathcal{I}_{k} = \left\langle \eta_{k}^{\dagger}\eta_{k}\right\rangle.
\end{equation} 
The total magnetization is not a conserved quantity for $\gamma\neq 0$, so relaxation occurs. Writing $\left\langle \cdot\right\rangle_{\mbox{\scriptsize gGE}}\,$ to denote the long-time average of an observable in the generalized Gibbs ensemble, we have
\begin{eqnarray}
& & \left\langle \hat{S}_{j}^{z}\right\rangle_{\mbox{\scriptsize gGE}}= \left\langle c_{j}^{\dagger}c_{j}\right\rangle_{\mbox{\scriptsize gGE}} - \frac{1}{2} = - \frac{1}{2} \nonumber\\ 
& & + \int_{-\pi}^{\pi}\frac{dk}{2\pi}\left[\cos^{2}\frac{\theta_{k}}{2}\left\langle \eta_{k}^{\dagger}\eta_{k}\right\rangle_{0} + \sin^{2}\frac{\theta_{k}}{2}\left\langle \eta_{-k}\eta_{-k}^{\dagger}\right\rangle_{0}\right].
\end{eqnarray}
Here $\left\langle \cdot\right\rangle_{0}\,$ is used to label an expectation value in the initial state. For the integrals of motion, $\left\langle \hat{\mathcal{I}}_{k}\right\rangle_{\mbox{\scriptsize gGE}} = \left\langle \hat{\mathcal{I}}_{k}\right\rangle_{0}$. To proceed, let us consider the left (right) side of a domain wall state with initial magnetization $m_{0}\,$ ($-m_{0}$). Locally, the ground state of $c$ fermions is a Fermi sea filled to momentum $k^{+} = \frac{\pi}{2} + \pi m_{0}$ ($k^{-} = \frac{\pi}{2} - \pi m_{0}$). In other words, each end is effectively its own semi-infinite subsystem with a symmetrically filled Fermi sea so we may write 
\begin{equation}
\left\langle \eta_{-k}\eta_{-k}^{\dagger}\right\rangle^{\pm}_{0} = 1 - \left\langle \eta_{k}^{\dagger}\eta_{k}\right\rangle_{0}^{\pm}.
\end{equation}
These initial expectation values are then
\begin{eqnarray}
\left\langle \eta_{k}^{\dagger}\eta_{k}\right\rangle_{0}^{\pm} & = & \cos^{2}\frac{\theta_{k}}{2}\left\langle c_{k}^{\dagger}c_{k}\right\rangle_{0}^{\pm} + \sin^{2}\frac{\theta_{k}}{2}\left\langle c_{-k}c_{-k}^{\dagger}\right\rangle_{0}^{\pm},\\
& = & \sin^{2}\frac{\theta_{k}}{2} + \cos\theta_{k}\Theta(k^{\pm}-|k|).
\end{eqnarray}
Taking $+\,$ ($-$) to formally represent the left (right) side, we have
\begin{eqnarray}
\left\langle \hat{S}_{j}^{z}\right\rangle^{\pm}_{\mbox{\scriptsize gGE}} & = & -\frac{1}{2} + \int_{-\pi}^{\pi}\frac{dk}{2\pi}\left[\frac{1}{2} - \frac{1}{2}\cos\theta_{k} \right. \nonumber\\
& +  & \left. \cos\theta_{k}\left\langle \eta_{k}^{\dagger}\eta_{k}\right\rangle^{\pm}_{0}\right],\\
& = & \int_{-k^{\pm}}^{k^{\pm}}\frac{dk}{2\pi}\cos^{2}\theta_{k} - \frac{1}{2}\int_{-\pi}^{\pi}\frac{dk}{2\pi}\cos^{2}\theta_{k}.
\end{eqnarray}
For $\theta_{k} = 0\,$ ($XX$ model), this trivially evaluates to $\pm m_{0}$. In the critical transverse-Ising model limit $\cos\theta_{k} = \left|\sin\frac{k}{2}\right|$, and we have
\begin{equation}
\left\langle \hat{S}_{j}\right\rangle^{\pm}_{\mbox{\scriptsize gGE}} = \pm \frac{m_{0}}{2} - \frac{\cos(\pi m_{0})}{2\pi},
\end{equation}
so that for a maximum jump in initial magnetization (from $\frac{1}{2}\,$ to $-\frac{1}{2}$), we have 
\begin{equation}
\left\langle \hat{S}_{j}\right\rangle^{\pm}_{\mbox{\scriptsize gGE}} = \pm \frac{1}{4}.
\end{equation}
For $m_{0} < \frac{1}{2}$, the sides relax {\it asymmetrically} so that the eventual equilibration of the central region results in non-zero magnetization when $h\neq 0$, as depicted in Fig.~\ref{fig:nessapproach}. In addition to the generalized Gibbs' ensemble prediction for the initial relaxation, we also depict the time-dependent behavior obtained from Eq.~(\ref{eq:cjt}) using the Eqs.~(\ref{eq:Fmt}) and (\ref{eq:Gmt}) for time evolution and the initial state presented in Ref.~\cite{Antal1999}, corresponding to a domain wall with height $m_{0} < \frac{1}{2}$. The $h=0\,$ limit turns out to be particularly interesting in terms of the gapped dynamics. The dispersion in Eq.~(\ref{eq:xydisp}) corresponds to a mode velocity
\begin{eqnarray}
v_{k} & = & \partial_{k}\epsilon_{k},\\
& = & \pm \frac{(1-\gamma^{2})\sin 2k - \frac{h}{J}\sin k}{2\epsilon_{k}},
\end{eqnarray}
which vanishes at $h = 0\,$ as $\gamma \rightarrow 1$. This vanishing mode velocity is reflected in the dynamics by keeping $h=0\,$ and letting $\gamma\,$ approach unity. The domain wall spreads very slowly for $\gamma \sim 1$. 
\begin{figure}
\begin{center}
\includegraphics[totalheight=6.25cm]{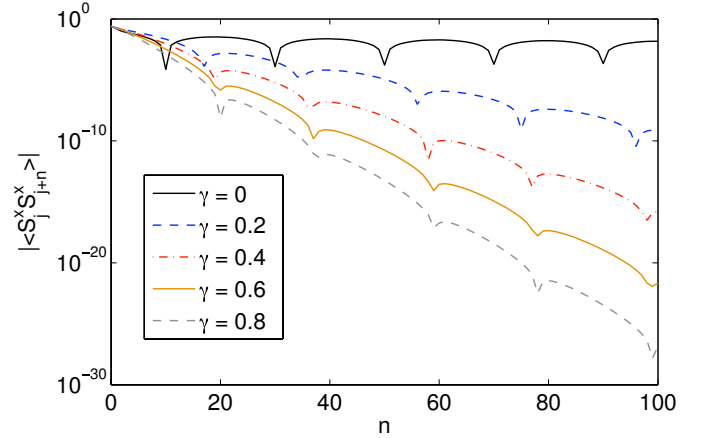}
\caption{(Color online) Magnitude of $\mathcal{C}^{xx}(n)\,$ in central non-equilibrium steady state for $m_{0} = 0.05$, $h = 0$, and various values of $\gamma$. Power-law decay is known to form the envelope of oscillations for $\gamma = 0\,$ whereas exponential decay sets in for increasing $\gamma$. While the current drops rapidly as $\gamma\,$ increases, the spatial oscillation wavelength is only weakly affected by the anisotropy. We note that for $h=0\,$ the magnetization in the central region approaches zero.}
\label{fig:cxxgam}
\end{center}
\end{figure}
In the limit of a very small domain wall $m_{0} \ll \frac{1}{2}$, the resulting current is determined by the value of $\cos\theta_{k}\,$ evaluated at the Fermi momentum $k = \frac{\pi}{2}$,
\begin{equation}
\left\langle \hat{\mathcal{J}}^{z}_{j}\right\rangle_{\mbox{\scriptsize NESS}} \xrightarrow{m_{0} \ll \frac{1}{2}} \frac{m_{0}h}{\sqrt{\frac{h^{2}}{J^{2}} + \gamma^{2}}} \;\;\;\; (0<\frac{h}{J},\gamma<1).\label{eq:smallmo}
\end{equation}
Though this limiting current depends strongly on $\gamma\,$ and $h\,$ in the for domain walls $m_{0} = \frac{1}{2}$, the oscillation wavelength in $\mathcal{C}^{xx}_{n}(t\rightarrow\infty)\,$ appears rigidly fixed by the magnetization jump [Eq.~(\ref{eq:cxxcos2})] and is insensitive to the details of $h\,$ and $\gamma$. For small $m_{0}$, Fig.~\ref{fig:cxxgam} depicts some variation in spatial oscillation wavelength as $\gamma\,$ is varied for small $m_{0}$. Note that Fig.~\ref{fig:cxxgam} does not depict the regime in which $m_{0}\,$ is small enough compared to unity for Eq.~(\ref{eq:smallmo}) to apply. Additionally, the case $h = 0\,$ should be treated separately, and Eq.~(\ref{eq:nesscur}) gives
\begin{equation}
\left\langle\hat{\mathcal{J}}^{z}_{j}\right\rangle_{\mbox{\scriptsize NESS}} =\frac{J}{\pi\left(1-\gamma^{2}\right)}\left(-\gamma + \sqrt{1-(1-\gamma^{2})\cos^{2}[\pi m_{0}]}\right)\label{eq:jness2} \;\;\;\; 
\end{equation}
which is valid for $0<m_{0}\leq \frac{1}{2}$, and reduces to Eq.~(\ref{eq:jupper}) as $m_{0}\rightarrow \frac{1}{2}$. The main result of this section is thus a strong dependence of the persistent current on the particular values of $h\,$ and $\gamma\,$ in the final Hamiltonian, for any value of $m_{0}$. By contrast, the spatial oscillations observed in $\mathcal{C}^{xx}_{n}\,$ are entirely fixed for maximal domain walls ($m_{0} = \frac{1}{2}$) and appear to show only a weak dependence on $\gamma$ compared to the significant variation in resulting current. 
\section{Conclusions}
\label{sec:concl}
In this paper, we have computed the long-time expectation values for observables after a quantum quench from the isotropic $XX$ model to the anisotropic $XY$ model. By applying an initial magnetic field gradient, the ground state of the initial Hamiltonian forms a spatially inhomogeneous domain-wall magnetization profile. In the critical transverse-Ising limit, we can obtain explicit expressions [Eqs.~(\ref{eq:bat})-(\ref{eq:bbt})], valid for all times. Of significant interest is the system's behavior in the long-time limit within a certain region of space over which the domain wall profile has relaxed to a spatially homogeneous value. We find an exponential envelope to the decay of transverse correlations with oscillations of a wavelength which is set by the magnetization jump in the initial state. While this oscillation wavelength is identical to that observed in the isotropic $XX$ model, the residual spin current is reduced due to the corresponding operator only having partial overlap with the conserved quantities in the anisotropic $XY$ model. This exponential decay is a consequence of beginning with a highly excited initial state with respect to the ground state of the transverse-Ising model, and we also demonstrate the emergence of exponential decay when performing a simple anisotropy quench from the ground state of the $XX$ model in the absence of external magnetic fields. It appears that the exponential decay, common to both homogeneous and inhomogeneous initial states created in the $XX$ model is a consequence of strong dephasing effects due to the abrupt change in the natural eigenbasis of the system \cite{Barthel2008}.

 In addition to obtaining the full dynamics in the critical transverse-Ising limit, we have extracted the long-time limit of basic observables for arbitrary anisotropy $\gamma$, magnetic field $h\,$ and domain-wall height $m_{0}$. Beginning from a current-carrying state in the $XX$ model, it is observed that a rapid quench to the $XY$ model for arbitrary $\gamma\,$ and $h\,$ results in a long-time steady state in which the spin current is equal to its initial value, despite this operator not being a conserved quantity of the final Hamiltonian. By contrast, if we consider a domain-wall initial state of height $m_{0}\,$ which would give rise to a current $\left\langle \mathcal{\hat{J}}^{z}_{j}\right\rangle = \frac{J}{\pi}\,$ in the long-time limit of evolution in the $XX$ model, it is shown that the corresponding long-time current is reduced from this value by an amount which depends on $\gamma\,$ and $h$. For $\gamma >0\,$ an energy gap exists in the system which affects the velocity of expansion of the domain wall, thereby affecting the current which saturates at a constant in the center of the system. We have presented explicit results for the steady-state current at long times in various regimes. Expressions such as Eq.~(\ref{eq:jness2}) which contain exact expressions valid for a wide range in parameter values (here, anisotropy $\gamma\,$ and domain wall height $m_{0}$) should prove useful especially in accounting for perturbative effects due to weak anisotropy. 

Recent experimental efforts have succeeded in observing details of far-from-equilibrium, one-dimensional systems with impressive precision, including power-law decay and phase oscillations of correlation functions \cite{Bloch2015}. A clear direction for future work is to determine how robust the results of this paper are with respect to small amounts of dissipation and weak interactions, inevitably present in any experimental situation. Numerical investigations which incorporate the effects of coupling anisotropy and interactions would be a valuable first step in this direction. While the $XXZ$ model is well studied for its nontrivial integrable interactions, significantly less attention is paid to the less symmetric $XYZ$ model, which is also integrable. It would interesting to explore the breaking of integrability through non-equilibrium situations such as those presented in this paper in the context of a fully interacting model such as the $XYZ$ spin chain with tunable next-nearest interactions as an integrability-breaking perturbation. Furthermore, the exact results presented in this work provide further benchmarks for testing continuum methods such as bosonization in out-of-equilibrium situations.

\acknowledgments
The author is grateful for support provided by the Joint School of Nanoscience and Nanoengineering during the completion of this work.

\appendix 
\section{Sums over products of Bessel functions}
\label{sec:bessel}
The sums in Eqs.~(\ref{eq:bat}) and (\ref{eq:bbt}) are of two basic forms,
\begin{eqnarray}
S_{k,l}^{(1)}(x) & = & \sum_{m=1}^{M}J_{k+2m}(x)J_{k+l+2m}(x),\label{eq:finitesum}\\
S_{k,l}^{(2)}(x) & = & \sum_{m>0}J_{k+2m}(x)J_{k+l+2m}(x).\label{eq:evensum}
\end{eqnarray}
Asymptotically, for $x \gg |k|,|l|$, the finite sums of the form in Eq.~(\ref{eq:finitesum}) make vanishing contributions, which decay as $\mathcal{O}(x^{-1})$. Taking the large-argument limit and replacing the Bessel functions by their asymptotic form,
\begin{equation}
J_{\nu}(x) \sim \sqrt{\frac{2}{\pi x}}\cos\left[x - \frac{\pi \nu}{2}-\frac{\pi}{4}\right],\label{eq:asymbessel}
\end{equation}
these sums are bounded
\begin{eqnarray}
\left|\sum_{m=1}^{M}J_{k+2m}(x)J_{k+l+2m}(x) \right| & < &  \frac{2M}{\pi x}\sim  \mathcal{O}(x^{-1}),
\end{eqnarray}
contributing to the region $|j|, |n| \gg Jt \gg 0\,$ after the initial global relaxation of the magnetization has occurred but before the domain wall has spread to the region containing site $j$. In the limit $|j|,|n| \ll Jt$, these sums in Eqs.~(\ref{eq:bat})-(\ref{eq:bbt}) vanish as $\mathcal{O}(t^{-1})$. The nonequilibrium steady state formed at long times is determined by the infinite sums of the form given in Eq.~(\ref{eq:evensum}). We may extract the infinite-time limit by appealing to a known sum rule \cite{Antal1999}
\begin{eqnarray}
\mathcal{T}_{k,l}(x) & \equiv & \sum_{m=1}^{\infty}J_{k+m}J_{k+l+m},\\
& = & \frac{x}{2l}\left[J_{k+1}J_{k+l}-J_{k}J_{k+l+1}\right]\label{eq:sumrule},
\end{eqnarray}
where the argument $x\,$ has been suppressed for brevity on the right-hand side. We may write
\begin{eqnarray}
\mathcal{T}_{k,l} & = & \sum_{m>0}^{\infty}\left[J_{k+2m}J_{k+l+2m} + J_{k+2m-1}J_{k+l+2m-1}\right],\\
& = &   S_{k,l}^{(2)} + S_{k-1,l}^{(2)}.
\end{eqnarray}
For $x \gg |k|, |l|\,$ we have $S_{k,l}^{(2)} \approx S_{k-1,l}^{(x)}\,$ in the homogeneous non-equilibrium steady state, so
\begin{equation}
\lim_{x\rightarrow\infty}\mathcal{T}_{k,l}(x) = 2\lim_{x\rightarrow\infty}S_{k,l}(x).
\end{equation}
Applying the asymptotic expansion Eq.~(\ref{eq:asymbessel}) to Eq.~(\ref{eq:sumrule}) we arrive at 
\begin{equation}
\lim_{x\rightarrow\infty}\sum_{m>0}J_{k+2m}J_{k+l+2m} = \frac{\sin\left(\frac{\pi l}{2}\right)}{2\pi l}.\label{eq:evensumresult}
\end{equation}
Substituting Eq.~(\ref{eq:evensumresult}) in Eqs.~(\ref{eq:bat})-(\ref{eq:bbt}) and neglecting contributions which vanish as $t^{-1}\,$ yields Eqs.~(\ref{eq:baness})-(\ref{eq:bbness}).

\section{Pfaffians for spin-spin correlation functions}
\label{sec:pfaffian}
It is known \cite{Barouch2,LSM1961} that the transverse two-point correlation functions are given by
\begin{eqnarray}
\left\langle \hat{S}_{j}^{x}\hat{S}_{j+n}^{x}\right\rangle & = & \frac{1}{4} \left\langle B_{j}A_{j+1}\cdots B_{j+n-1}A_{j+n}\right\rangle,\label{eq:cxxwick}\\
\left\langle \hat{S}_{j}^{y}\hat{S}_{j+n}^{y}\right\rangle & = & \frac{1}{4}(-1)^{n}\left\langle A_{j}B_{j+1}\cdots A_{j+n-1}B_{j+n}\right\rangle\label{eq:cyywick}.
\end{eqnarray}
The Pfaffian of an antisymmetric matrix $A_{ij} = -A_{ji}\,$ is the square root of the matrix determinant,
\begin{equation}
\left[\mbox{pf}A\right]^{2} = \mbox{det}A.
\end{equation}
Applying Wick's theorem, Eqs.~(\ref{eq:cxxwick}) and (\ref{eq:cyywick}) are readily expressed as Pfaffians \cite{Stolze1992},
\begin{widetext}
\begin{eqnarray}
 \left\langle B_{j}A_{j+1}B_{j+1}\cdots A_{j+n-1}B_{j+n-1}A_{j+n}\right\rangle & = & \left|\begin{array}{ccccc} \left\langle B_{j}A_{j+1}\right\rangle &  \left\langle B_{j}B_{j+1}\right\rangle & \cdots &  \left\langle B_{j}B_{j+n-1}\right\rangle & \left\langle B_{j}A_{j+n}\right\rangle\\  & \left\langle A_{j+1}B_{j+1}\right\rangle & \cdots & \left\langle A_{j+1}B_{j+n-1}\right\rangle & \left\langle A_{j+1}A_{j+n}\right\rangle\\ & & \ddots & \vdots & \vdots \\ & & & \left\langle A_{j+n-1}B_{j+n-1}\right\rangle & \left\langle A_{j+n-1}A_{j+n}\right\rangle\\ & & & & \left\langle B_{j+n-1}A_{j+n}\right\rangle\end{array}\right|,\label{eq:cxxpfaf}\\
\left\langle A_{j}B_{j+1}A_{j+1}\cdots B_{j+n-1}A_{j+n-1}B_{j+n}\right\rangle & = & \left|\begin{array}{ccccc} \left\langle A_{j}B_{j+1}\right\rangle &  \left\langle A_{j}A_{j+1}\right\rangle & \cdots&\left\langle A_{j}B_{j+n-1}\right\rangle  & \left\langle A_{j}B_{j+n}\right\rangle\\  & \left\langle B_{j+1}A_{j+1}\right\rangle & \cdots & \left\langle B_{j+1}A_{j+n-1}\right\rangle & \left\langle B_{j+1}B_{j+n}\right\rangle\\ & & \ddots & \vdots & \vdots \\ & & & \left\langle B_{j+n-1}A_{j+n-1}\right\rangle & \left\langle B_{j+n-1}B_{j+n}\right\rangle\\ & & & & \left\langle A_{j+n-1}B_{j+n}\right\rangle\end{array}\right|.\label{eq:cyypfaf}
\end{eqnarray}
\end{widetext}
Barouch and collaborators \cite{Barouch2} use a somewhat re-arranged form for Eq.~(\ref{eq:cxxpfaf}), making its equivalence to Eq.~(\ref{eq:cxxness}) most transparent. The evaluation of these Pfaffians may be accomplished by taking the square root of the determinant of the corresponding skew-symmetric matrix \cite{Lancaster2010a} or through direct expansion in minors. An immediate drawback to the former approach is an indeterminacy in the sign of the correlation, since its square is being computed directly. Fortunately, efficient libraries for direct Pfaffian computation are available in most common programming languages \cite{Wimmer2012}. It is straightforward to verify that $\mathcal{C}^{yy}_{\mbox{\scriptsize NESS}}(n) = \mathcal{C}^{xx}_{\mbox{\scriptsize NESS}}(n)\,$ for those cases noted in this paper by numerically evaluating Eqs.~(\ref{eq:cxxpfaf}) and (\ref{eq:cyypfaf}). To this end, we require the expressions,
\begin{eqnarray}
\left\langle B_{j}(t)A_{j+n}(t)\right\rangle & = & 2\mbox{Re}\left[\left\langle c_{j}^{\dagger}c_{j+n}\right\rangle\right] - 2\mbox{Re}\left[\left\langle c_{j}c_{j+n}\right\rangle\right]\nonumber\\
 & - &  \delta_{n=0},\\
\left\langle A_{j}(t)B_{j+n}(t)\right\rangle & = & -2\mbox{Re}\left[\left\langle c_{j}^{\dagger}c_{j+n}\right\rangle\right] - 2\mbox{Re}\left[\left\langle c_{j}c_{j+n}\right\rangle\right]\nonumber\\
 & +&  \delta_{n=0},\\
\left\langle A_{j}(t)A_{j+n}(t)\right\rangle & = & 2i\mbox{Im}\left[\left\langle c_{j}^{\dagger}c_{j+n}\right\rangle\right]+ 2i\mbox{Im}\left[\left\langle c_{j}c_{j+n}\right\rangle\right]\nonumber\\
 & + &  \delta_{n=0},\\
\left\langle B_{j}(t)B_{j+n}(t)\right\rangle & = & -2i\mbox{Im}\left[\left\langle c_{j}^{\dagger}c_{j+n}\right\rangle\right] + 2i\mbox{Im}\left[\left\langle c_{j}c_{j+n}\right\rangle\right]\nonumber\\
 & - &  \delta_{n=0}.
\end{eqnarray}
\bibliography{xybib} 

\begin{thebibliography}{71}%
\makeatletter
\providecommand \@ifxundefined [1]{%
 \@ifx{#1\undefined}
}%
\providecommand \@ifnum [1]{%
 \ifnum #1\expandafter \@firstoftwo
 \else \expandafter \@secondoftwo
 \fi
}%
\providecommand \@ifx [1]{%
 \ifx #1\expandafter \@firstoftwo
 \else \expandafter \@secondoftwo
 \fi
}%
\providecommand \natexlab [1]{#1}%
\providecommand \enquote  [1]{``#1''}%
\providecommand \bibnamefont  [1]{#1}%
\providecommand \bibfnamefont [1]{#1}%
\providecommand \citenamefont [1]{#1}%
\providecommand \href@noop [0]{\@secondoftwo}%
\providecommand \href [0]{\begingroup \@sanitize@url \@href}%
\providecommand \@href[1]{\@@startlink{#1}\@@href}%
\providecommand \@@href[1]{\endgroup#1\@@endlink}%
\providecommand \@sanitize@url [0]{\catcode `\\12\catcode `\$12\catcode
  `\&12\catcode `\#12\catcode `\^12\catcode `\_12\catcode `\%12\relax}%
\providecommand \@@startlink[1]{}%
\providecommand \@@endlink[0]{}%
\providecommand \url  [0]{\begingroup\@sanitize@url \@url }%
\providecommand \@url [1]{\endgroup\@href {#1}{\urlprefix }}%
\providecommand \urlprefix  [0]{URL }%
\providecommand \Eprint [0]{\href }%
\providecommand \doibase [0]{http://dx.doi.org/}%
\providecommand \selectlanguage [0]{\@gobble}%
\providecommand \bibinfo  [0]{\@secondoftwo}%
\providecommand \bibfield  [0]{\@secondoftwo}%
\providecommand \translation [1]{[#1]}%
\providecommand \BibitemOpen [0]{}%
\providecommand \bibitemStop [0]{}%
\providecommand \bibitemNoStop [0]{.\EOS\space}%
\providecommand \EOS [0]{\spacefactor3000\relax}%
\providecommand \BibitemShut  [1]{\csname bibitem#1\endcsname}%
\let\auto@bib@innerbib\@empty
\bibitem [{\citenamefont {Polkovnikov}\ \emph {et~al.}(2011)\citenamefont
  {Polkovnikov}, \citenamefont {Sengupta}, \citenamefont {Silva},\ and\
  \citenamefont {Vengalattore}}]{Silva2011}%
  \BibitemOpen
  \bibfield  {author} {\bibinfo {author} {\bibfnamefont {A.}~\bibnamefont
  {Polkovnikov}}, \bibinfo {author} {\bibfnamefont {K.}~\bibnamefont
  {Sengupta}}, \bibinfo {author} {\bibfnamefont {A.}~\bibnamefont {Silva}}, \
  and\ \bibinfo {author} {\bibfnamefont {M.}~\bibnamefont {Vengalattore}},\
  }\href {\doibase 10.1103/RevModPhys.83.863} {\bibfield  {journal} {\bibinfo
  {journal} {Rev. Mod. Phys.}\ }\textbf {\bibinfo {volume} {83}},\ \bibinfo
  {pages} {863} (\bibinfo {year} {2011})}\BibitemShut {NoStop}%
\bibitem [{\citenamefont {Cazalilla}\ \emph {et~al.}(2011)\citenamefont
  {Cazalilla}, \citenamefont {Citro}, \citenamefont {Giamarchi}, \citenamefont
  {Orignac},\ and\ \citenamefont {Rigol}}]{Cazalilla2011}%
  \BibitemOpen
  \bibfield  {author} {\bibinfo {author} {\bibfnamefont {M.~A.}\ \bibnamefont
  {Cazalilla}}, \bibinfo {author} {\bibfnamefont {R.}~\bibnamefont {Citro}},
  \bibinfo {author} {\bibfnamefont {T.}~\bibnamefont {Giamarchi}}, \bibinfo
  {author} {\bibfnamefont {E.}~\bibnamefont {Orignac}}, \ and\ \bibinfo
  {author} {\bibfnamefont {M.}~\bibnamefont {Rigol}},\ }\href {\doibase
  10.1103/RevModPhys.83.1405} {\bibfield  {journal} {\bibinfo  {journal} {Rev.
  Mod. Phys.}\ }\textbf {\bibinfo {volume} {83}},\ \bibinfo {pages} {1405}
  (\bibinfo {year} {2011})}\BibitemShut {NoStop}%
\bibitem [{\citenamefont {Eisert}\ \emph {et~al.}(2015)\citenamefont {Eisert},
  \citenamefont {Friesdorf},\ and\ \citenamefont {Gogolin}}]{Eisert2015}%
  \BibitemOpen
  \bibfield  {author} {\bibinfo {author} {\bibfnamefont {J.}~\bibnamefont
  {Eisert}}, \bibinfo {author} {\bibfnamefont {M.}~\bibnamefont {Friesdorf}}, \
  and\ \bibinfo {author} {\bibfnamefont {C.}~\bibnamefont {Gogolin}},\
  }\href@noop {} {\bibfield  {journal} {\bibinfo  {journal} {Nat. Phys.}\
  }\textbf {\bibinfo {volume} {11}},\ \bibinfo {pages} {124} (\bibinfo {year}
  {2015})}\BibitemShut {NoStop}%
\bibitem [{\citenamefont {Feynman}(1982)}]{Feynman1982}%
  \BibitemOpen
  \bibfield  {author} {\bibinfo {author} {\bibfnamefont {R.~P.}\ \bibnamefont
  {Feynman}},\ }\href@noop {} {\bibfield  {journal} {\bibinfo  {journal} {Int.
  J. Theor. Phys.}\ }\textbf {\bibinfo {volume} {21}},\ \bibinfo {pages} {467 }
  (\bibinfo {year} {1982})}\BibitemShut {NoStop}%
\bibitem [{\citenamefont {Chin}\ \emph {et~al.}(2010)\citenamefont {Chin},
  \citenamefont {Grimm}, \citenamefont {Julienne},\ and\ \citenamefont
  {Tiesinga}}]{Chin2010}%
  \BibitemOpen
  \bibfield  {author} {\bibinfo {author} {\bibfnamefont {C.}~\bibnamefont
  {Chin}}, \bibinfo {author} {\bibfnamefont {R.}~\bibnamefont {Grimm}},
  \bibinfo {author} {\bibfnamefont {P.}~\bibnamefont {Julienne}}, \ and\
  \bibinfo {author} {\bibfnamefont {E.}~\bibnamefont {Tiesinga}},\ }\href@noop
  {} {\bibfield  {journal} {\bibinfo  {journal} {Rev. Mod. Phys.}\ }\textbf
  {\bibinfo {volume} {82}},\ \bibinfo {pages} {1225} (\bibinfo {year}
  {2010})}\BibitemShut {NoStop}%
\bibitem [{\citenamefont {Vidmar}\ \emph
  {et~al.}(2015{\natexlab{a}})\citenamefont {Vidmar}, \citenamefont
  {Ronzheimer}, \citenamefont {Schreiber}, \citenamefont {Braun}, \citenamefont
  {Hodgman}, \citenamefont {Langer}, \citenamefont {Heidrich-Meisner},
  \citenamefont {Bloch},\ and\ \citenamefont {Schneider}}]{Bloch2015}%
  \BibitemOpen
  \bibfield  {author} {\bibinfo {author} {\bibfnamefont {L.}~\bibnamefont
  {Vidmar}}, \bibinfo {author} {\bibfnamefont {J.~P.}\ \bibnamefont
  {Ronzheimer}}, \bibinfo {author} {\bibfnamefont {M.}~\bibnamefont
  {Schreiber}}, \bibinfo {author} {\bibfnamefont {S.}~\bibnamefont {Braun}},
  \bibinfo {author} {\bibfnamefont {S.~S.}\ \bibnamefont {Hodgman}}, \bibinfo
  {author} {\bibfnamefont {S.}~\bibnamefont {Langer}}, \bibinfo {author}
  {\bibfnamefont {F.}~\bibnamefont {Heidrich-Meisner}}, \bibinfo {author}
  {\bibfnamefont {I.}~\bibnamefont {Bloch}}, \ and\ \bibinfo {author}
  {\bibfnamefont {U.}~\bibnamefont {Schneider}},\ }\href {\doibase
  10.1103/PhysRevLett.115.175301} {\bibfield  {journal} {\bibinfo  {journal}
  {Phys. Rev. Lett.}\ }\textbf {\bibinfo {volume} {115}},\ \bibinfo {pages}
  {175301} (\bibinfo {year} {2015}{\natexlab{a}})}\BibitemShut {NoStop}%
\bibitem [{\citenamefont {Senko}\ \emph {et~al.}(2015)\citenamefont {Senko},
  \citenamefont {Richerme}, \citenamefont {Smith}, \citenamefont {Lee},
  \citenamefont {Cohen}, \citenamefont {Retzker},\ and\ \citenamefont
  {Monroe}}]{Senko2015}%
  \BibitemOpen
  \bibfield  {author} {\bibinfo {author} {\bibfnamefont {C.}~\bibnamefont
  {Senko}}, \bibinfo {author} {\bibfnamefont {P.}~\bibnamefont {Richerme}},
  \bibinfo {author} {\bibfnamefont {J.}~\bibnamefont {Smith}}, \bibinfo
  {author} {\bibfnamefont {A.}~\bibnamefont {Lee}}, \bibinfo {author}
  {\bibfnamefont {I.}~\bibnamefont {Cohen}}, \bibinfo {author} {\bibfnamefont
  {A.}~\bibnamefont {Retzker}}, \ and\ \bibinfo {author} {\bibfnamefont
  {C.}~\bibnamefont {Monroe}},\ }\href {\doibase 10.1103/PhysRevX.5.021026}
  {\bibfield  {journal} {\bibinfo  {journal} {Phys. Rev. X}\ }\textbf {\bibinfo
  {volume} {5}},\ \bibinfo {pages} {021026} (\bibinfo {year}
  {2015})}\BibitemShut {NoStop}%
\bibitem [{\citenamefont {Kinoshita}\ \emph {et~al.}(2006)\citenamefont
  {Kinoshita}, \citenamefont {Wenger},\ and\ \citenamefont
  {Weiss}}]{Weiss2006}%
  \BibitemOpen
  \bibfield  {author} {\bibinfo {author} {\bibfnamefont {T.}~\bibnamefont
  {Kinoshita}}, \bibinfo {author} {\bibfnamefont {T.}~\bibnamefont {Wenger}}, \
  and\ \bibinfo {author} {\bibfnamefont {D.~S.}\ \bibnamefont {Weiss}},\ }\href
  {\doibase 10.1038/nature04693} {\bibfield  {journal} {\bibinfo  {journal}
  {Nature}\ }\textbf {\bibinfo {volume} {440}},\ \bibinfo {pages} {900}
  (\bibinfo {year} {2006})}\BibitemShut {NoStop}%
\bibitem [{\citenamefont {Rigol}\ \emph {et~al.}(2006)\citenamefont {Rigol},
  \citenamefont {Muramatsu},\ and\ \citenamefont {Olshanii}}]{Rigol2006}%
  \BibitemOpen
  \bibfield  {author} {\bibinfo {author} {\bibfnamefont {M.}~\bibnamefont
  {Rigol}}, \bibinfo {author} {\bibfnamefont {A.}~\bibnamefont {Muramatsu}}, \
  and\ \bibinfo {author} {\bibfnamefont {M.}~\bibnamefont {Olshanii}},\ }\href
  {\doibase 10.1103/PhysRevA.74.053616} {\bibfield  {journal} {\bibinfo
  {journal} {Phys. Rev. A}\ }\textbf {\bibinfo {volume} {74}},\ \bibinfo
  {pages} {053616} (\bibinfo {year} {2006})}\BibitemShut {NoStop}%
\bibitem [{\citenamefont {Rigol}\ \emph {et~al.}(2007)\citenamefont {Rigol},
  \citenamefont {Dunjko}, \citenamefont {Yurovsky},\ and\ \citenamefont
  {Olshanii}}]{Rigol2007}%
  \BibitemOpen
  \bibfield  {author} {\bibinfo {author} {\bibfnamefont {M.}~\bibnamefont
  {Rigol}}, \bibinfo {author} {\bibfnamefont {V.}~\bibnamefont {Dunjko}},
  \bibinfo {author} {\bibfnamefont {V.}~\bibnamefont {Yurovsky}}, \ and\
  \bibinfo {author} {\bibfnamefont {M.}~\bibnamefont {Olshanii}},\ }\href@noop
  {} {\bibfield  {journal} {\bibinfo  {journal} {Phys. Rev. Lett.}\ }\textbf
  {\bibinfo {volume} {98}},\ \bibinfo {pages} {050405} (\bibinfo {year}
  {2007})}\BibitemShut {NoStop}%
\bibitem [{\citenamefont {Rigol}\ \emph {et~al.}(2008)\citenamefont {Rigol},
  \citenamefont {Dunjko},\ and\ \citenamefont {Olshanii}}]{Rigol2008}%
  \BibitemOpen
  \bibfield  {author} {\bibinfo {author} {\bibfnamefont {M.}~\bibnamefont
  {Rigol}}, \bibinfo {author} {\bibfnamefont {V.}~\bibnamefont {Dunjko}}, \
  and\ \bibinfo {author} {\bibfnamefont {M.}~\bibnamefont {Olshanii}},\
  }\href@noop {} {\bibfield  {journal} {\bibinfo  {journal} {Nature}\ }\textbf
  {\bibinfo {volume} {452}},\ \bibinfo {pages} {854} (\bibinfo {year}
  {2008})}\BibitemShut {NoStop}%
\bibitem [{\citenamefont {Cazalilla}(2006)}]{Cazalilla2006}%
  \BibitemOpen
  \bibfield  {author} {\bibinfo {author} {\bibfnamefont {M.~A.}\ \bibnamefont
  {Cazalilla}},\ }\href {\doibase 10.1103/PhysRevLett.97.156403} {\bibfield
  {journal} {\bibinfo  {journal} {Phys. Rev. Lett.}\ }\textbf {\bibinfo
  {volume} {97}},\ \bibinfo {pages} {156403} (\bibinfo {year}
  {2006})}\BibitemShut {NoStop}%
\bibitem [{\citenamefont {Kollar}\ \emph {et~al.}(2011)\citenamefont {Kollar},
  \citenamefont {Wolf},\ and\ \citenamefont {Eckstein}}]{Kollar2011}%
  \BibitemOpen
  \bibfield  {author} {\bibinfo {author} {\bibfnamefont {M.}~\bibnamefont
  {Kollar}}, \bibinfo {author} {\bibfnamefont {F.~A.}\ \bibnamefont {Wolf}}, \
  and\ \bibinfo {author} {\bibfnamefont {M.}~\bibnamefont {Eckstein}},\ }\href
  {\doibase 10.1103/PhysRevB.84.054304} {\bibfield  {journal} {\bibinfo
  {journal} {Phys. Rev. B}\ }\textbf {\bibinfo {volume} {84}},\ \bibinfo
  {pages} {054304} (\bibinfo {year} {2011})}\BibitemShut {NoStop}%
\bibitem [{\citenamefont {Marcuzzi}\ \emph {et~al.}(2013)\citenamefont
  {Marcuzzi}, \citenamefont {Marino}, \citenamefont {Gambassi},\ and\
  \citenamefont {Silva}}]{Marcuzzi2013}%
  \BibitemOpen
  \bibfield  {author} {\bibinfo {author} {\bibfnamefont {M.}~\bibnamefont
  {Marcuzzi}}, \bibinfo {author} {\bibfnamefont {J.}~\bibnamefont {Marino}},
  \bibinfo {author} {\bibfnamefont {A.}~\bibnamefont {Gambassi}}, \ and\
  \bibinfo {author} {\bibfnamefont {A.}~\bibnamefont {Silva}},\ }\href
  {\doibase 10.1103/PhysRevLett.111.197203} {\bibfield  {journal} {\bibinfo
  {journal} {Phys. Rev. Lett.}\ }\textbf {\bibinfo {volume} {111}},\ \bibinfo
  {pages} {197203} (\bibinfo {year} {2013})}\BibitemShut {NoStop}%
\bibitem [{\citenamefont {Langen}\ \emph {et~al.}(2015)\citenamefont {Langen},
  \citenamefont {Erne}, \citenamefont {Geiger}, \citenamefont {Rauer},
  \citenamefont {Schweigler}, \citenamefont {Kuhnert}, \citenamefont
  {Rohringer}, \citenamefont {Mazets}, \citenamefont {Gasenzer},\ and\
  \citenamefont {Schmiedmayer}}]{Langen2015}%
  \BibitemOpen
  \bibfield  {author} {\bibinfo {author} {\bibfnamefont {T.}~\bibnamefont
  {Langen}}, \bibinfo {author} {\bibfnamefont {S.}~\bibnamefont {Erne}},
  \bibinfo {author} {\bibfnamefont {R.}~\bibnamefont {Geiger}}, \bibinfo
  {author} {\bibfnamefont {B.}~\bibnamefont {Rauer}}, \bibinfo {author}
  {\bibfnamefont {T.}~\bibnamefont {Schweigler}}, \bibinfo {author}
  {\bibfnamefont {M.}~\bibnamefont {Kuhnert}}, \bibinfo {author} {\bibfnamefont
  {W.}~\bibnamefont {Rohringer}}, \bibinfo {author} {\bibfnamefont {I.~E.}\
  \bibnamefont {Mazets}}, \bibinfo {author} {\bibfnamefont {T.}~\bibnamefont
  {Gasenzer}}, \ and\ \bibinfo {author} {\bibfnamefont {J.}~\bibnamefont
  {Schmiedmayer}},\ }\href@noop {} {\bibfield  {journal} {\bibinfo  {journal}
  {Science}\ }\textbf {\bibinfo {volume} {348}},\ \bibinfo {pages} {207}
  (\bibinfo {year} {2015})}\BibitemShut {NoStop}%
\bibitem [{\citenamefont {Bortolin}\ and\ \citenamefont
  {Iucci}(2015)}]{Bortolin2015}%
  \BibitemOpen
  \bibfield  {author} {\bibinfo {author} {\bibfnamefont {T.~S.}\ \bibnamefont
  {Bortolin}}\ and\ \bibinfo {author} {\bibfnamefont {A.}~\bibnamefont
  {Iucci}},\ }\href {\doibase 10.1103/PhysRevB.91.024301} {\bibfield  {journal}
  {\bibinfo  {journal} {Phys. Rev. B}\ }\textbf {\bibinfo {volume} {91}},\
  \bibinfo {pages} {024301} (\bibinfo {year} {2015})}\BibitemShut {NoStop}%
\bibitem [{\citenamefont {Gobert}\ \emph {et~al.}(2005)\citenamefont {Gobert},
  \citenamefont {Kollath}, \citenamefont {Schollw\"ock},\ and\ \citenamefont
  {Sch\"utz}}]{Schollwock2005}%
  \BibitemOpen
  \bibfield  {author} {\bibinfo {author} {\bibfnamefont {D.}~\bibnamefont
  {Gobert}}, \bibinfo {author} {\bibfnamefont {C.}~\bibnamefont {Kollath}},
  \bibinfo {author} {\bibfnamefont {U.}~\bibnamefont {Schollw\"ock}}, \ and\
  \bibinfo {author} {\bibfnamefont {G.}~\bibnamefont {Sch\"utz}},\ }\href
  {\doibase 10.1103/PhysRevE.71.036102} {\bibfield  {journal} {\bibinfo
  {journal} {Phys. Rev. E}\ }\textbf {\bibinfo {volume} {71}},\ \bibinfo
  {pages} {036102} (\bibinfo {year} {2005})}\BibitemShut {NoStop}%
\bibitem [{\citenamefont {Lancaster}\ and\ \citenamefont
  {Mitra}(2010)}]{Lancaster2010a}%
  \BibitemOpen
  \bibfield  {author} {\bibinfo {author} {\bibfnamefont {J.}~\bibnamefont
  {Lancaster}}\ and\ \bibinfo {author} {\bibfnamefont {A.}~\bibnamefont
  {Mitra}},\ }\href {\doibase 10.1103/PhysRevE.81.061134} {\bibfield  {journal}
  {\bibinfo  {journal} {Phys. Rev. E}\ }\textbf {\bibinfo {volume} {81}},\
  \bibinfo {pages} {061134} (\bibinfo {year} {2010})}\BibitemShut {NoStop}%
\bibitem [{\citenamefont {Foster}\ \emph {et~al.}(2010)\citenamefont {Foster},
  \citenamefont {Yuzbashyan},\ and\ \citenamefont {Altshuler}}]{Foster2010}%
  \BibitemOpen
  \bibfield  {author} {\bibinfo {author} {\bibfnamefont {M.~S.}\ \bibnamefont
  {Foster}}, \bibinfo {author} {\bibfnamefont {E.~A.}\ \bibnamefont
  {Yuzbashyan}}, \ and\ \bibinfo {author} {\bibfnamefont {B.~L.}\ \bibnamefont
  {Altshuler}},\ }\href {\doibase 10.1103/PhysRevLett.105.135701} {\bibfield
  {journal} {\bibinfo  {journal} {Phys. Rev. Lett.}\ }\textbf {\bibinfo
  {volume} {105}},\ \bibinfo {pages} {135701} (\bibinfo {year}
  {2010})}\BibitemShut {NoStop}%
\bibitem [{\citenamefont {Foster}\ \emph {et~al.}(2011)\citenamefont {Foster},
  \citenamefont {Berkelbach}, \citenamefont {Reichman},\ and\ \citenamefont
  {Yuzbashyan}}]{Foster2011}%
  \BibitemOpen
  \bibfield  {author} {\bibinfo {author} {\bibfnamefont {M.~S.}\ \bibnamefont
  {Foster}}, \bibinfo {author} {\bibfnamefont {T.~C.}\ \bibnamefont
  {Berkelbach}}, \bibinfo {author} {\bibfnamefont {D.~R.}\ \bibnamefont
  {Reichman}}, \ and\ \bibinfo {author} {\bibfnamefont {E.~A.}\ \bibnamefont
  {Yuzbashyan}},\ }\href {\doibase 10.1103/PhysRevB.84.085146} {\bibfield
  {journal} {\bibinfo  {journal} {Phys. Rev. B}\ }\textbf {\bibinfo {volume}
  {84}},\ \bibinfo {pages} {085146} (\bibinfo {year} {2011})}\BibitemShut
  {NoStop}%
\bibitem [{\citenamefont {Sabetta}\ and\ \citenamefont
  {Misguich}(2013)}]{Sabetta2013}%
  \BibitemOpen
  \bibfield  {author} {\bibinfo {author} {\bibfnamefont {T.}~\bibnamefont
  {Sabetta}}\ and\ \bibinfo {author} {\bibfnamefont {G.}~\bibnamefont
  {Misguich}},\ }\href {\doibase 10.1103/PhysRevB.88.245114} {\bibfield
  {journal} {\bibinfo  {journal} {Phys. Rev. B}\ }\textbf {\bibinfo {volume}
  {88}},\ \bibinfo {pages} {245114} (\bibinfo {year} {2013})}\BibitemShut
  {NoStop}%
\bibitem [{\citenamefont {Bravo}\ \emph {et~al.}(2013)\citenamefont {Bravo},
  \citenamefont {Dobry}, \citenamefont {Mastrogiuseppe},\ and\ \citenamefont
  {Gazza}}]{Bravo2013}%
  \BibitemOpen
  \bibfield  {author} {\bibinfo {author} {\bibfnamefont {B.}~\bibnamefont
  {Bravo}}, \bibinfo {author} {\bibfnamefont {A.}~\bibnamefont {Dobry}},
  \bibinfo {author} {\bibfnamefont {D.}~\bibnamefont {Mastrogiuseppe}}, \ and\
  \bibinfo {author} {\bibfnamefont {C.}~\bibnamefont {Gazza}},\ }\href@noop {}
  {\bibfield  {journal} {\bibinfo  {journal} {Phys. Rev. B}\ }\textbf {\bibinfo
  {volume} {88}},\ \bibinfo {pages} {195125} (\bibinfo {year}
  {2013})}\BibitemShut {NoStop}%
\bibitem [{\citenamefont {Mitra}\ and\ \citenamefont
  {Giamarchi}(2011)}]{Mitra2011}%
  \BibitemOpen
  \bibfield  {author} {\bibinfo {author} {\bibfnamefont {A.}~\bibnamefont
  {Mitra}}\ and\ \bibinfo {author} {\bibfnamefont {T.}~\bibnamefont
  {Giamarchi}},\ }\href {\doibase 10.1103/PhysRevLett.107.150602} {\bibfield
  {journal} {\bibinfo  {journal} {Phys. Rev. Lett.}\ }\textbf {\bibinfo
  {volume} {107}},\ \bibinfo {pages} {150602} (\bibinfo {year}
  {2011})}\BibitemShut {NoStop}%
\bibitem [{\citenamefont {Mitra}\ and\ \citenamefont
  {Giamarchi}(2012)}]{Mitra2012}%
  \BibitemOpen
  \bibfield  {author} {\bibinfo {author} {\bibfnamefont {A.}~\bibnamefont
  {Mitra}}\ and\ \bibinfo {author} {\bibfnamefont {T.}~\bibnamefont
  {Giamarchi}},\ }\href {\doibase 10.1103/PhysRevB.85.075117} {\bibfield
  {journal} {\bibinfo  {journal} {Phys. Rev. B}\ }\textbf {\bibinfo {volume}
  {85}},\ \bibinfo {pages} {075117} (\bibinfo {year} {2012})}\BibitemShut
  {NoStop}%
\bibitem [{\citenamefont {Tavora}\ and\ \citenamefont
  {Mitra}(2013)}]{Tavora2013}%
  \BibitemOpen
  \bibfield  {author} {\bibinfo {author} {\bibfnamefont {M.}~\bibnamefont
  {Tavora}}\ and\ \bibinfo {author} {\bibfnamefont {A.}~\bibnamefont {Mitra}},\
  }\href {\doibase 10.1103/PhysRevB.88.115144} {\bibfield  {journal} {\bibinfo
  {journal} {Phys. Rev. B}\ }\textbf {\bibinfo {volume} {88}},\ \bibinfo
  {pages} {115144} (\bibinfo {year} {2013})}\BibitemShut {NoStop}%
\bibitem [{\citenamefont {Cazalilla}\ \emph {et~al.}(2012)\citenamefont
  {Cazalilla}, \citenamefont {Iucci},\ and\ \citenamefont
  {Chung}}]{Cazalilla2012}%
  \BibitemOpen
  \bibfield  {author} {\bibinfo {author} {\bibfnamefont {M.~A.}\ \bibnamefont
  {Cazalilla}}, \bibinfo {author} {\bibfnamefont {A.}~\bibnamefont {Iucci}}, \
  and\ \bibinfo {author} {\bibfnamefont {M.-C.}\ \bibnamefont {Chung}},\
  }\href@noop {} {\bibfield  {journal} {\bibinfo  {journal} {Phys. Rev. E}\
  }\textbf {\bibinfo {volume} {85}},\ \bibinfo {pages} {011133} (\bibinfo
  {year} {2012})}\BibitemShut {NoStop}%
\bibitem [{\citenamefont {Weld}\ \emph {et~al.}(2009)\citenamefont {Weld},
  \citenamefont {Medley}, \citenamefont {Miyake}, \citenamefont {Hucul},
  \citenamefont {Pritchard},\ and\ \citenamefont {Ketterle}}]{Weld2009}%
  \BibitemOpen
  \bibfield  {author} {\bibinfo {author} {\bibfnamefont {D.~M.}\ \bibnamefont
  {Weld}}, \bibinfo {author} {\bibfnamefont {P.}~\bibnamefont {Medley}},
  \bibinfo {author} {\bibfnamefont {H.}~\bibnamefont {Miyake}}, \bibinfo
  {author} {\bibfnamefont {D.}~\bibnamefont {Hucul}}, \bibinfo {author}
  {\bibfnamefont {D.~E.}\ \bibnamefont {Pritchard}}, \ and\ \bibinfo {author}
  {\bibfnamefont {W.}~\bibnamefont {Ketterle}},\ }\href@noop {} {\bibfield
  {journal} {\bibinfo  {journal} {Phys. Rev. Lett.}\ }\textbf {\bibinfo
  {volume} {103}},\ \bibinfo {pages} {245301} (\bibinfo {year}
  {2009})}\BibitemShut {NoStop}%
\bibitem [{\citenamefont {Lieb}\ \emph {et~al.}(1961)\citenamefont {Lieb},
  \citenamefont {Schultz},\ and\ \citenamefont {Mattis}}]{LSM1961}%
  \BibitemOpen
  \bibfield  {author} {\bibinfo {author} {\bibfnamefont {E.}~\bibnamefont
  {Lieb}}, \bibinfo {author} {\bibfnamefont {T.}~\bibnamefont {Schultz}}, \
  and\ \bibinfo {author} {\bibfnamefont {D.}~\bibnamefont {Mattis}},\ }\href
  {\doibase http://dx.doi.org/10.1016/0003-4916(61)90115-4} {\bibfield
  {journal} {\bibinfo  {journal} {Annals of Physics}\ }\textbf {\bibinfo
  {volume} {16}},\ \bibinfo {pages} {407 } (\bibinfo {year}
  {1961})}\BibitemShut {NoStop}%
\bibitem [{\citenamefont {Barouch}\ \emph {et~al.}(1970)\citenamefont
  {Barouch}, \citenamefont {McCoy},\ and\ \citenamefont {Dresden}}]{Barouch1}%
  \BibitemOpen
  \bibfield  {author} {\bibinfo {author} {\bibfnamefont {E.}~\bibnamefont
  {Barouch}}, \bibinfo {author} {\bibfnamefont {B.~M.}\ \bibnamefont {McCoy}},
  \ and\ \bibinfo {author} {\bibfnamefont {M.}~\bibnamefont {Dresden}},\ }\href
  {\doibase 10.1103/PhysRevA.2.1075} {\bibfield  {journal} {\bibinfo  {journal}
  {Phys. Rev. A}\ }\textbf {\bibinfo {volume} {2}},\ \bibinfo {pages} {1075}
  (\bibinfo {year} {1970})}\BibitemShut {NoStop}%
\bibitem [{\citenamefont {Barouch}\ and\ \citenamefont
  {McCoy}(1971{\natexlab{a}})}]{Barouch2}%
  \BibitemOpen
  \bibfield  {author} {\bibinfo {author} {\bibfnamefont {E.}~\bibnamefont
  {Barouch}}\ and\ \bibinfo {author} {\bibfnamefont {B.~M.}\ \bibnamefont
  {McCoy}},\ }\href {\doibase 10.1103/PhysRevA.3.786} {\bibfield  {journal}
  {\bibinfo  {journal} {Phys. Rev. A}\ }\textbf {\bibinfo {volume} {3}},\
  \bibinfo {pages} {786} (\bibinfo {year} {1971}{\natexlab{a}})}\BibitemShut
  {NoStop}%
\bibitem [{\citenamefont {Pfeuty}(1970)}]{Pfeuty1970}%
  \BibitemOpen
  \bibfield  {author} {\bibinfo {author} {\bibfnamefont {P.}~\bibnamefont
  {Pfeuty}},\ }\href {\doibase http://dx.doi.org/10.1016/0003-4916(70)90270-8}
  {\bibfield  {journal} {\bibinfo  {journal} {Annals of Physics}\ }\textbf
  {\bibinfo {volume} {57}},\ \bibinfo {pages} {79 } (\bibinfo {year}
  {1970})}\BibitemShut {NoStop}%
\bibitem [{\citenamefont {Antal}\ \emph {et~al.}(1997)\citenamefont {Antal},
  \citenamefont {R\'acz},\ and\ \citenamefont {Sasv\'ari}}]{Antal1997}%
  \BibitemOpen
  \bibfield  {author} {\bibinfo {author} {\bibfnamefont {T.}~\bibnamefont
  {Antal}}, \bibinfo {author} {\bibfnamefont {Z.}~\bibnamefont {R\'acz}}, \
  and\ \bibinfo {author} {\bibfnamefont {L.}~\bibnamefont {Sasv\'ari}},\ }\href
  {\doibase 10.1103/PhysRevLett.78.167} {\bibfield  {journal} {\bibinfo
  {journal} {Phys. Rev. Lett.}\ }\textbf {\bibinfo {volume} {78}},\ \bibinfo
  {pages} {167} (\bibinfo {year} {1997})}\BibitemShut {NoStop}%
\bibitem [{\citenamefont {Antal}\ \emph {et~al.}(1998)\citenamefont {Antal},
  \citenamefont {R\'acz}, \citenamefont {R\'akos},\ and\ \citenamefont
  {Sch\"utz}}]{Antal1998}%
  \BibitemOpen
  \bibfield  {author} {\bibinfo {author} {\bibfnamefont {T.}~\bibnamefont
  {Antal}}, \bibinfo {author} {\bibfnamefont {Z.}~\bibnamefont {R\'acz}},
  \bibinfo {author} {\bibfnamefont {A.}~\bibnamefont {R\'akos}}, \ and\
  \bibinfo {author} {\bibfnamefont {G.~M.}\ \bibnamefont {Sch\"utz}},\ }\href
  {\doibase 10.1103/PhysRevE.57.5184} {\bibfield  {journal} {\bibinfo
  {journal} {Phys. Rev. E}\ }\textbf {\bibinfo {volume} {57}},\ \bibinfo
  {pages} {5184} (\bibinfo {year} {1998})}\BibitemShut {NoStop}%
\bibitem [{\citenamefont {Antal}\ \emph {et~al.}(1999)\citenamefont {Antal},
  \citenamefont {R\'acz}, \citenamefont {R\'akos},\ and\ \citenamefont
  {Sch\"utz}}]{Antal1999}%
  \BibitemOpen
  \bibfield  {author} {\bibinfo {author} {\bibfnamefont {T.}~\bibnamefont
  {Antal}}, \bibinfo {author} {\bibfnamefont {Z.}~\bibnamefont {R\'acz}},
  \bibinfo {author} {\bibfnamefont {A.}~\bibnamefont {R\'akos}}, \ and\
  \bibinfo {author} {\bibfnamefont {G.~M.}\ \bibnamefont {Sch\"utz}},\ }\href
  {\doibase 10.1103/PhysRevE.59.4912} {\bibfield  {journal} {\bibinfo
  {journal} {Phys. Rev. E}\ }\textbf {\bibinfo {volume} {59}},\ \bibinfo
  {pages} {4912} (\bibinfo {year} {1999})}\BibitemShut {NoStop}%
\bibitem [{\citenamefont {Karevski}(2002)}]{Karevski2002}%
  \BibitemOpen
  \bibfield  {author} {\bibinfo {author} {\bibfnamefont {D.}~\bibnamefont
  {Karevski}},\ }\href {\doibase 10.1140/epjb/e20020139} {\bibfield  {journal}
  {\bibinfo  {journal} {Eur. Phys. J. B}\ }\textbf {\bibinfo {volume} {27}},\
  \bibinfo {pages} {147} (\bibinfo {year} {2002})}\BibitemShut {NoStop}%
\bibitem [{\citenamefont {Rossini}\ \emph {et~al.}(2010)\citenamefont
  {Rossini}, \citenamefont {Suzuki}, \citenamefont {Mussardo}, \citenamefont
  {Santoro},\ and\ \citenamefont {Silva}}]{Rossini2011}%
  \BibitemOpen
  \bibfield  {author} {\bibinfo {author} {\bibfnamefont {D.}~\bibnamefont
  {Rossini}}, \bibinfo {author} {\bibfnamefont {S.}~\bibnamefont {Suzuki}},
  \bibinfo {author} {\bibfnamefont {G.}~\bibnamefont {Mussardo}}, \bibinfo
  {author} {\bibfnamefont {G.~E.}\ \bibnamefont {Santoro}}, \ and\ \bibinfo
  {author} {\bibfnamefont {A.}~\bibnamefont {Silva}},\ }\href {\doibase
  10.1103/PhysRevB.82.144302} {\bibfield  {journal} {\bibinfo  {journal} {Phys.
  Rev. B}\ }\textbf {\bibinfo {volume} {82}},\ \bibinfo {pages} {144302}
  (\bibinfo {year} {2010})}\BibitemShut {NoStop}%
\bibitem [{\citenamefont {Caneva}\ \emph {et~al.}(2011)\citenamefont {Caneva},
  \citenamefont {Canovi}, \citenamefont {Rossini}, \citenamefont {Santoro},\
  and\ \citenamefont {Silva}}]{Caneva2011}%
  \BibitemOpen
  \bibfield  {author} {\bibinfo {author} {\bibfnamefont {T.}~\bibnamefont
  {Caneva}}, \bibinfo {author} {\bibfnamefont {E.}~\bibnamefont {Canovi}},
  \bibinfo {author} {\bibfnamefont {D.}~\bibnamefont {Rossini}}, \bibinfo
  {author} {\bibfnamefont {G.~E.}\ \bibnamefont {Santoro}}, \ and\ \bibinfo
  {author} {\bibfnamefont {A.}~\bibnamefont {Silva}},\ }\href@noop {}
  {\bibfield  {journal} {\bibinfo  {journal} {Journal of Statistical Mechanics:
  Theory and Experiment}\ }\textbf {\bibinfo {volume} {2011}},\ \bibinfo
  {pages} {P07015} (\bibinfo {year} {2011})}\BibitemShut {NoStop}%
\bibitem [{\citenamefont {Alba}\ and\ \citenamefont
  {Heidrich-Meisner}(2014)}]{Alba2014}%
  \BibitemOpen
  \bibfield  {author} {\bibinfo {author} {\bibfnamefont {V.}~\bibnamefont
  {Alba}}\ and\ \bibinfo {author} {\bibfnamefont {F.}~\bibnamefont
  {Heidrich-Meisner}},\ }\href@noop {} {\bibfield  {journal} {\bibinfo
  {journal} {Phys. Rev. B}\ }\textbf {\bibinfo {volume} {90}},\ \bibinfo
  {pages} {075144} (\bibinfo {year} {2014})}\BibitemShut {NoStop}%
\bibitem [{\citenamefont {Mukherjee}\ \emph {et~al.}(2007)\citenamefont
  {Mukherjee}, \citenamefont {Divakaran}, \citenamefont {Dutta},\ and\
  \citenamefont {Sen}}]{Mukherjee2007}%
  \BibitemOpen
  \bibfield  {author} {\bibinfo {author} {\bibfnamefont {V.}~\bibnamefont
  {Mukherjee}}, \bibinfo {author} {\bibfnamefont {U.}~\bibnamefont
  {Divakaran}}, \bibinfo {author} {\bibfnamefont {A.}~\bibnamefont {Dutta}}, \
  and\ \bibinfo {author} {\bibfnamefont {D.}~\bibnamefont {Sen}},\ }\href@noop
  {} {\bibfield  {journal} {\bibinfo  {journal} {Phys. Rev. B}\ }\textbf
  {\bibinfo {volume} {76}},\ \bibinfo {pages} {174303} (\bibinfo {year}
  {2007})}\BibitemShut {NoStop}%
\bibitem [{\citenamefont {Mukherjee}\ \emph {et~al.}(2008)\citenamefont
  {Mukherjee}, \citenamefont {Dutta},\ and\ \citenamefont
  {Sen}}]{Mukherjee2008}%
  \BibitemOpen
  \bibfield  {author} {\bibinfo {author} {\bibfnamefont {V.}~\bibnamefont
  {Mukherjee}}, \bibinfo {author} {\bibfnamefont {A.}~\bibnamefont {Dutta}}, \
  and\ \bibinfo {author} {\bibfnamefont {D.}~\bibnamefont {Sen}},\ }\href@noop
  {} {\bibfield  {journal} {\bibinfo  {journal} {Phys. Rev. B}\ }\textbf
  {\bibinfo {volume} {77}},\ \bibinfo {pages} {214427} (\bibinfo {year}
  {2008})}\BibitemShut {NoStop}%
\bibitem [{\citenamefont {Deng}\ \emph {et~al.}(2011)\citenamefont {Deng},
  \citenamefont {Ortiz},\ and\ \citenamefont {Viola}}]{Deng2011}%
  \BibitemOpen
  \bibfield  {author} {\bibinfo {author} {\bibfnamefont {S.}~\bibnamefont
  {Deng}}, \bibinfo {author} {\bibfnamefont {G.}~\bibnamefont {Ortiz}}, \ and\
  \bibinfo {author} {\bibfnamefont {L.}~\bibnamefont {Viola}},\ }\href@noop {}
  {\bibfield  {journal} {\bibinfo  {journal} {Phys. Rev. B}\ }\textbf {\bibinfo
  {volume} {83}},\ \bibinfo {pages} {094304} (\bibinfo {year}
  {2011})}\BibitemShut {NoStop}%
\bibitem [{\citenamefont {Calabrese}\ and\ \citenamefont
  {Cardy}(2006)}]{Cardy2006}%
  \BibitemOpen
  \bibfield  {author} {\bibinfo {author} {\bibfnamefont {P.}~\bibnamefont
  {Calabrese}}\ and\ \bibinfo {author} {\bibfnamefont {J.}~\bibnamefont
  {Cardy}},\ }\href {\doibase 10.1103/PhysRevLett.96.136801} {\bibfield
  {journal} {\bibinfo  {journal} {Phys. Rev. Lett.}\ }\textbf {\bibinfo
  {volume} {96}},\ \bibinfo {pages} {136801} (\bibinfo {year}
  {2006})}\BibitemShut {NoStop}%
\bibitem [{\citenamefont {Sotiriadis}\ and\ \citenamefont
  {Cardy}(2008)}]{Cardy2008}%
  \BibitemOpen
  \bibfield  {author} {\bibinfo {author} {\bibfnamefont {S.}~\bibnamefont
  {Sotiriadis}}\ and\ \bibinfo {author} {\bibfnamefont {J.}~\bibnamefont
  {Cardy}},\ }\href@noop {} {\bibfield  {journal} {\bibinfo  {journal} {Journal
  of Statistical Mechanics: Theory and Experiment}\ }\textbf {\bibinfo {volume}
  {2008}},\ \bibinfo {pages} {P11003} (\bibinfo {year} {2008})}\BibitemShut
  {NoStop}%
\bibitem [{\citenamefont {Calabrese}\ \emph {et~al.}(2008)\citenamefont
  {Calabrese}, \citenamefont {Hagendorf},\ and\ \citenamefont
  {Doussal}}]{Calabrese2008}%
  \BibitemOpen
  \bibfield  {author} {\bibinfo {author} {\bibfnamefont {P.}~\bibnamefont
  {Calabrese}}, \bibinfo {author} {\bibfnamefont {C.}~\bibnamefont
  {Hagendorf}}, \ and\ \bibinfo {author} {\bibfnamefont {P.~L.}\ \bibnamefont
  {Doussal}},\ }\href@noop {} {\bibfield  {journal} {\bibinfo  {journal}
  {Journal of Statistical Mechanics: Theory and Experiment}\ }\textbf {\bibinfo
  {volume} {2008}},\ \bibinfo {pages} {P07013} (\bibinfo {year}
  {2008})}\BibitemShut {NoStop}%
\bibitem [{\citenamefont {Bayocboc}\ and\ \citenamefont
  {Paraan}(2015)}]{Bayocboc2015}%
  \BibitemOpen
  \bibfield  {author} {\bibinfo {author} {\bibfnamefont {F.~A.}\ \bibnamefont
  {Bayocboc}}\ and\ \bibinfo {author} {\bibfnamefont {F.~N.~C.}\ \bibnamefont
  {Paraan}},\ }\href@noop {} {\bibfield  {journal} {\bibinfo  {journal} {Phys.
  Rev. E}\ }\textbf {\bibinfo {volume} {92}},\ \bibinfo {pages} {032142}
  (\bibinfo {year} {2015})}\BibitemShut {NoStop}%
\bibitem [{\citenamefont {Viehmann}\ \emph {et~al.}(2013)\citenamefont
  {Viehmann}, \citenamefont {von Delft},\ and\ \citenamefont
  {Marquardt}}]{Viehmann2013}%
  \BibitemOpen
  \bibfield  {author} {\bibinfo {author} {\bibfnamefont {O.}~\bibnamefont
  {Viehmann}}, \bibinfo {author} {\bibfnamefont {J.}~\bibnamefont {von Delft}},
  \ and\ \bibinfo {author} {\bibfnamefont {F.}~\bibnamefont {Marquardt}},\
  }\href {\doibase 10.1103/PhysRevLett.110.030601} {\bibfield  {journal}
  {\bibinfo  {journal} {Phys. Rev. Lett.}\ }\textbf {\bibinfo {volume} {110}},\
  \bibinfo {pages} {030601} (\bibinfo {year} {2013})}\BibitemShut {NoStop}%
\bibitem [{\citenamefont {Heyl}\ \emph {et~al.}(2013)\citenamefont {Heyl},
  \citenamefont {Polkovnikov},\ and\ \citenamefont {Kehrein}}]{Heyl2013}%
  \BibitemOpen
  \bibfield  {author} {\bibinfo {author} {\bibfnamefont {M.}~\bibnamefont
  {Heyl}}, \bibinfo {author} {\bibfnamefont {A.}~\bibnamefont {Polkovnikov}}, \
  and\ \bibinfo {author} {\bibfnamefont {S.}~\bibnamefont {Kehrein}},\ }\href
  {\doibase 10.1103/PhysRevLett.110.135704} {\bibfield  {journal} {\bibinfo
  {journal} {Phys. Rev. Lett.}\ }\textbf {\bibinfo {volume} {110}},\ \bibinfo
  {pages} {135704} (\bibinfo {year} {2013})}\BibitemShut {NoStop}%
\bibitem [{\citenamefont {Kriel}\ \emph {et~al.}(2014)\citenamefont {Kriel},
  \citenamefont {Karrasch},\ and\ \citenamefont {Kehrein}}]{Kriel2014}%
  \BibitemOpen
  \bibfield  {author} {\bibinfo {author} {\bibfnamefont {J.~N.}\ \bibnamefont
  {Kriel}}, \bibinfo {author} {\bibfnamefont {C.}~\bibnamefont {Karrasch}}, \
  and\ \bibinfo {author} {\bibfnamefont {S.}~\bibnamefont {Kehrein}},\ }\href
  {\doibase 10.1103/PhysRevB.90.125106} {\bibfield  {journal} {\bibinfo
  {journal} {Phys. Rev. B}\ }\textbf {\bibinfo {volume} {90}},\ \bibinfo
  {pages} {125106} (\bibinfo {year} {2014})}\BibitemShut {NoStop}%
\bibitem [{\citenamefont {Zamolodchikov}(1989)}]{Zamolodchikov1989}%
  \BibitemOpen
  \bibfield  {author} {\bibinfo {author} {\bibfnamefont {A.~B.}\ \bibnamefont
  {Zamolodchikov}},\ }\href@noop {} {\bibfield  {journal} {\bibinfo  {journal}
  {Int. J. Mod. Phys. A}\ }\textbf {\bibinfo {volume} {04}},\ \bibinfo {pages}
  {4235} (\bibinfo {year} {1989})}\BibitemShut {NoStop}%
\bibitem [{\citenamefont {{Mussardo}}(2009)}]{Mussardo}%
  \BibitemOpen
  \bibfield  {author} {\bibinfo {author} {\bibfnamefont {G.}~\bibnamefont
  {{Mussardo}}},\ }\href@noop {} {\emph {\bibinfo {title} {Statistical Field
  Theory: An Introduction to Exactly Solved Models in Statistical Physics}}}\
  (\bibinfo  {publisher} {Oxford U Press},\ \bibinfo {address} {Oxford, UK},\
  \bibinfo {year} {2009})\BibitemShut {NoStop}%
\bibitem [{\citenamefont {Coldea}\ \emph {et~al.}(2010)\citenamefont {Coldea},
  \citenamefont {Tennant}, \citenamefont {Wheeler}, \citenamefont {Wawrzynska},
  \citenamefont {Prabhakaran}, \citenamefont {Telling}, \citenamefont
  {Habicht}, \citenamefont {Smeibidl},\ and\ \citenamefont
  {Kiefer}}]{Coldea2010}%
  \BibitemOpen
  \bibfield  {author} {\bibinfo {author} {\bibfnamefont {R.}~\bibnamefont
  {Coldea}}, \bibinfo {author} {\bibfnamefont {D.~A.}\ \bibnamefont {Tennant}},
  \bibinfo {author} {\bibfnamefont {E.~M.}\ \bibnamefont {Wheeler}}, \bibinfo
  {author} {\bibfnamefont {E.}~\bibnamefont {Wawrzynska}}, \bibinfo {author}
  {\bibfnamefont {D.}~\bibnamefont {Prabhakaran}}, \bibinfo {author}
  {\bibfnamefont {M.}~\bibnamefont {Telling}}, \bibinfo {author} {\bibfnamefont
  {K.}~\bibnamefont {Habicht}}, \bibinfo {author} {\bibfnamefont
  {P.}~\bibnamefont {Smeibidl}}, \ and\ \bibinfo {author} {\bibfnamefont
  {K.}~\bibnamefont {Kiefer}},\ }\href {\doibase 10.1126/science.1180085}
  {\bibfield  {journal} {\bibinfo  {journal} {Science}\ }\textbf {\bibinfo
  {volume} {327}},\ \bibinfo {pages} {177} (\bibinfo {year}
  {2010})}\BibitemShut {NoStop}%
\bibitem [{\citenamefont {Richerme}\ \emph {et~al.}(2014)\citenamefont
  {Richerme}, \citenamefont {Gong}, \citenamefont {Lee}, \citenamefont {Senko},
  \citenamefont {Smith}, \citenamefont {Foss-Feig}, \citenamefont {Michalakis},
  \citenamefont {Gorshkov},\ and\ \citenamefont {Monroe}}]{Richerme2014}%
  \BibitemOpen
  \bibfield  {author} {\bibinfo {author} {\bibfnamefont {P.}~\bibnamefont
  {Richerme}}, \bibinfo {author} {\bibfnamefont {Z.-X.}\ \bibnamefont {Gong}},
  \bibinfo {author} {\bibfnamefont {A.}~\bibnamefont {Lee}}, \bibinfo {author}
  {\bibfnamefont {C.}~\bibnamefont {Senko}}, \bibinfo {author} {\bibfnamefont
  {J.}~\bibnamefont {Smith}}, \bibinfo {author} {\bibfnamefont
  {M.}~\bibnamefont {Foss-Feig}}, \bibinfo {author} {\bibfnamefont
  {S.}~\bibnamefont {Michalakis}}, \bibinfo {author} {\bibfnamefont {A.~V.}\
  \bibnamefont {Gorshkov}}, \ and\ \bibinfo {author} {\bibfnamefont
  {C.}~\bibnamefont {Monroe}},\ }\href@noop {} {\bibfield  {journal} {\bibinfo
  {journal} {Nature}\ }\textbf {\bibinfo {volume} {511}},\ \bibinfo {pages}
  {198} (\bibinfo {year} {2014})}\BibitemShut {NoStop}%
\bibitem [{\citenamefont {Jurcevic}\ \emph {et~al.}(2015)\citenamefont
  {Jurcevic}, \citenamefont {Hauke}, \citenamefont {Maier}, \citenamefont
  {Hempel}, \citenamefont {Lanyon}, \citenamefont {Blatt},\ and\ \citenamefont
  {Roos}}]{Jurcevic2015}%
  \BibitemOpen
  \bibfield  {author} {\bibinfo {author} {\bibfnamefont {P.}~\bibnamefont
  {Jurcevic}}, \bibinfo {author} {\bibfnamefont {P.}~\bibnamefont {Hauke}},
  \bibinfo {author} {\bibfnamefont {C.}~\bibnamefont {Maier}}, \bibinfo
  {author} {\bibfnamefont {C.}~\bibnamefont {Hempel}}, \bibinfo {author}
  {\bibfnamefont {B.~P.}\ \bibnamefont {Lanyon}}, \bibinfo {author}
  {\bibfnamefont {R.}~\bibnamefont {Blatt}}, \ and\ \bibinfo {author}
  {\bibfnamefont {C.~F.}\ \bibnamefont {Roos}},\ }\href {\doibase
  10.1103/PhysRevLett.115.100501} {\bibfield  {journal} {\bibinfo  {journal}
  {Phys. Rev. Lett.}\ }\textbf {\bibinfo {volume} {115}},\ \bibinfo {pages}
  {100501} (\bibinfo {year} {2015})}\BibitemShut {NoStop}%
\bibitem [{\citenamefont {Calabrese}\ \emph {et~al.}(2011)\citenamefont
  {Calabrese}, \citenamefont {Essler},\ and\ \citenamefont
  {Fagotti}}]{Calabrese2011}%
  \BibitemOpen
  \bibfield  {author} {\bibinfo {author} {\bibfnamefont {P.}~\bibnamefont
  {Calabrese}}, \bibinfo {author} {\bibfnamefont {F.~H.~L.}\ \bibnamefont
  {Essler}}, \ and\ \bibinfo {author} {\bibfnamefont {M.}~\bibnamefont
  {Fagotti}},\ }\href {\doibase 10.1103/PhysRevLett.106.227203} {\bibfield
  {journal} {\bibinfo  {journal} {Phys. Rev. Lett.}\ }\textbf {\bibinfo
  {volume} {106}},\ \bibinfo {pages} {227203} (\bibinfo {year}
  {2011})}\BibitemShut {NoStop}%
\bibitem [{\citenamefont {Foini}\ \emph {et~al.}(2011)\citenamefont {Foini},
  \citenamefont {Cugliandolo},\ and\ \citenamefont {Gambassi}}]{Foini2011}%
  \BibitemOpen
  \bibfield  {author} {\bibinfo {author} {\bibfnamefont {L.}~\bibnamefont
  {Foini}}, \bibinfo {author} {\bibfnamefont {L.~F.}\ \bibnamefont
  {Cugliandolo}}, \ and\ \bibinfo {author} {\bibfnamefont {A.}~\bibnamefont
  {Gambassi}},\ }\href {\doibase 10.1103/PhysRevB.84.212404} {\bibfield
  {journal} {\bibinfo  {journal} {Phys. Rev. B}\ }\textbf {\bibinfo {volume}
  {84}},\ \bibinfo {pages} {212404} (\bibinfo {year} {2011})}\BibitemShut
  {NoStop}%
\bibitem [{\citenamefont {Subrahmanyam}(2003)}]{Subrahmanyam2003}%
  \BibitemOpen
  \bibfield  {author} {\bibinfo {author} {\bibfnamefont {V.}~\bibnamefont
  {Subrahmanyam}},\ }\href {\doibase 10.1103/PhysRevB.68.212407} {\bibfield
  {journal} {\bibinfo  {journal} {Phys. Rev. B}\ }\textbf {\bibinfo {volume}
  {68}},\ \bibinfo {pages} {212407} (\bibinfo {year} {2003})}\BibitemShut
  {NoStop}%
\bibitem [{\citenamefont {Rigol}\ and\ \citenamefont
  {Muramatsu}(2004)}]{Rigol2004}%
  \BibitemOpen
  \bibfield  {author} {\bibinfo {author} {\bibfnamefont {M.}~\bibnamefont
  {Rigol}}\ and\ \bibinfo {author} {\bibfnamefont {A.}~\bibnamefont
  {Muramatsu}},\ }\href {\doibase 10.1103/PhysRevLett.93.230404} {\bibfield
  {journal} {\bibinfo  {journal} {Phys. Rev. Lett.}\ }\textbf {\bibinfo
  {volume} {93}},\ \bibinfo {pages} {230404} (\bibinfo {year}
  {2004})}\BibitemShut {NoStop}%
\bibitem [{\citenamefont {Lancaster}\ \emph {et~al.}(2010)\citenamefont
  {Lancaster}, \citenamefont {Gull},\ and\ \citenamefont
  {Mitra}}]{Lancaster2010b}%
  \BibitemOpen
  \bibfield  {author} {\bibinfo {author} {\bibfnamefont {J.}~\bibnamefont
  {Lancaster}}, \bibinfo {author} {\bibfnamefont {E.}~\bibnamefont {Gull}}, \
  and\ \bibinfo {author} {\bibfnamefont {A.}~\bibnamefont {Mitra}},\ }\href
  {\doibase 10.1103/PhysRevB.82.235124} {\bibfield  {journal} {\bibinfo
  {journal} {Phys. Rev. B}\ }\textbf {\bibinfo {volume} {82}},\ \bibinfo
  {pages} {235124} (\bibinfo {year} {2010})}\BibitemShut {NoStop}%
\bibitem [{\citenamefont {Barouch}\ and\ \citenamefont
  {McCoy}(1971{\natexlab{b}})}]{Barouch3}%
  \BibitemOpen
  \bibfield  {author} {\bibinfo {author} {\bibfnamefont {E.}~\bibnamefont
  {Barouch}}\ and\ \bibinfo {author} {\bibfnamefont {B.~M.}\ \bibnamefont
  {McCoy}},\ }\href@noop {} {\bibfield  {journal} {\bibinfo  {journal} {Phys.
  Rev. A}\ }\textbf {\bibinfo {volume} {3}},\ \bibinfo {pages} {2137} (\bibinfo
  {year} {1971}{\natexlab{b}})}\BibitemShut {NoStop}%
\bibitem [{\citenamefont {McCoy}\ \emph {et~al.}(1971)\citenamefont {McCoy},
  \citenamefont {Barouch},\ and\ \citenamefont {Abraham}}]{Barouch4}%
  \BibitemOpen
  \bibfield  {author} {\bibinfo {author} {\bibfnamefont {B.~M.}\ \bibnamefont
  {McCoy}}, \bibinfo {author} {\bibfnamefont {E.}~\bibnamefont {Barouch}}, \
  and\ \bibinfo {author} {\bibfnamefont {D.~B.}\ \bibnamefont {Abraham}},\
  }\href {\doibase 10.1103/PhysRevA.4.2331} {\bibfield  {journal} {\bibinfo
  {journal} {Phys. Rev. A}\ }\textbf {\bibinfo {volume} {4}},\ \bibinfo {pages}
  {2331} (\bibinfo {year} {1971})}\BibitemShut {NoStop}%
\bibitem [{\citenamefont {Eisler}\ \emph {et~al.}(2009)\citenamefont {Eisler},
  \citenamefont {Iglói},\ and\ \citenamefont {Peschel}}]{Eisler2009}%
  \BibitemOpen
  \bibfield  {author} {\bibinfo {author} {\bibfnamefont {V.}~\bibnamefont
  {Eisler}}, \bibinfo {author} {\bibfnamefont {F.}~\bibnamefont {Iglói}}, \
  and\ \bibinfo {author} {\bibfnamefont {I.}~\bibnamefont {Peschel}},\
  }\href@noop {} {\bibfield  {journal} {\bibinfo  {journal} {Journal of
  Statistical Mechanics: Theory and Experiment}\ }\textbf {\bibinfo {volume}
  {2009}},\ \bibinfo {pages} {P02011} (\bibinfo {year} {2009})}\BibitemShut
  {NoStop}%
\bibitem [{\citenamefont {Deift}\ \emph {et~al.}(2011)\citenamefont {Deift},
  \citenamefont {Its},\ and\ \citenamefont {Krasovsky}}]{FH1}%
  \BibitemOpen
  \bibfield  {author} {\bibinfo {author} {\bibfnamefont {P.}~\bibnamefont
  {Deift}}, \bibinfo {author} {\bibfnamefont {A.}~\bibnamefont {Its}}, \ and\
  \bibinfo {author} {\bibfnamefont {I.}~\bibnamefont {Krasovsky}},\ }\href@noop
  {} {\bibfield  {journal} {\bibinfo  {journal} {Annals of Math.}\ }\textbf
  {\bibinfo {volume} {174}},\ \bibinfo {pages} {1243} (\bibinfo {year}
  {2011})}\BibitemShut {NoStop}%
\bibitem [{\citenamefont {Its}\ and\ \citenamefont {Korepin}(2009)}]{FH2}%
  \BibitemOpen
  \bibfield  {author} {\bibinfo {author} {\bibfnamefont {A.~R.}\ \bibnamefont
  {Its}}\ and\ \bibinfo {author} {\bibfnamefont {V.~E.}\ \bibnamefont
  {Korepin}},\ }\href@noop {} {\bibfield  {journal} {\bibinfo  {journal} {J.
  Stat. Phys.}\ }\textbf {\bibinfo {volume} {137}},\ \bibinfo {pages} {1014}
  (\bibinfo {year} {2009})}\BibitemShut {NoStop}%
\bibitem [{\citenamefont {Basor}\ and\ \citenamefont {Morrison}(1994)}]{FH4}%
  \BibitemOpen
  \bibfield  {author} {\bibinfo {author} {\bibfnamefont {E.~L.}\ \bibnamefont
  {Basor}}\ and\ \bibinfo {author} {\bibfnamefont {K.~E.}\ \bibnamefont
  {Morrison}},\ }\href@noop {} {\bibfield  {journal} {\bibinfo  {journal}
  {Linear Algebra Appl.}\ }\textbf {\bibinfo {volume} {202}},\ \bibinfo {pages}
  {129} (\bibinfo {year} {1994})}\BibitemShut {NoStop}%
\bibitem [{\citenamefont {Ovchinnikov}(2007)}]{Ovchinnikov2007}%
  \BibitemOpen
  \bibfield  {author} {\bibinfo {author} {\bibfnamefont {A.}~\bibnamefont
  {Ovchinnikov}},\ }\href@noop {} {\bibfield  {journal} {\bibinfo  {journal}
  {Physics Letters A}\ }\textbf {\bibinfo {volume} {366}},\ \bibinfo {pages}
  {357 } (\bibinfo {year} {2007})}\BibitemShut {NoStop}%
\bibitem [{\citenamefont {Sengupta}\ \emph {et~al.}(2004)\citenamefont
  {Sengupta}, \citenamefont {Powell},\ and\ \citenamefont
  {Sachdev}}]{Sengupta2004}%
  \BibitemOpen
  \bibfield  {author} {\bibinfo {author} {\bibfnamefont {K.}~\bibnamefont
  {Sengupta}}, \bibinfo {author} {\bibfnamefont {S.}~\bibnamefont {Powell}}, \
  and\ \bibinfo {author} {\bibfnamefont {S.}~\bibnamefont {Sachdev}},\ }\href
  {\doibase 10.1103/PhysRevA.69.053616} {\bibfield  {journal} {\bibinfo
  {journal} {Phys. Rev. A}\ }\textbf {\bibinfo {volume} {69}},\ \bibinfo
  {pages} {053616} (\bibinfo {year} {2004})}\BibitemShut {NoStop}%
\bibitem [{\citenamefont {Vidmar}\ \emph
  {et~al.}(2015{\natexlab{b}})\citenamefont {Vidmar}, \citenamefont {Iyer},\
  and\ \citenamefont {Rigol}}]{Vidmar2015}%
  \BibitemOpen
  \bibfield  {author} {\bibinfo {author} {\bibfnamefont {L.}~\bibnamefont
  {Vidmar}}, \bibinfo {author} {\bibfnamefont {D.}~\bibnamefont {Iyer}}, \ and\
  \bibinfo {author} {\bibfnamefont {M.}~\bibnamefont {Rigol}},\ }\href@noop {}
  {\bibfield  {journal} {\bibinfo  {journal} {(unpublished) arXiv:1512.05373}\
  } (\bibinfo {year} {2015}{\natexlab{b}})}\BibitemShut {NoStop}%
\bibitem [{\citenamefont {Iucci}\ and\ \citenamefont
  {Cazalilla}(2009)}]{Iucci2009}%
  \BibitemOpen
  \bibfield  {author} {\bibinfo {author} {\bibfnamefont {A.}~\bibnamefont
  {Iucci}}\ and\ \bibinfo {author} {\bibfnamefont {M.~A.}\ \bibnamefont
  {Cazalilla}},\ }\href {\doibase 10.1103/PhysRevA.80.063619} {\bibfield
  {journal} {\bibinfo  {journal} {Phys. Rev. A}\ }\textbf {\bibinfo {volume}
  {80}},\ \bibinfo {pages} {063619} (\bibinfo {year} {2009})}\BibitemShut
  {NoStop}%
\bibitem [{\citenamefont {Barthel}\ and\ \citenamefont
  {Schollw\"ock}(2008)}]{Barthel2008}%
  \BibitemOpen
  \bibfield  {author} {\bibinfo {author} {\bibfnamefont {T.}~\bibnamefont
  {Barthel}}\ and\ \bibinfo {author} {\bibfnamefont {U.}~\bibnamefont
  {Schollw\"ock}},\ }\href {\doibase 10.1103/PhysRevLett.100.100601} {\bibfield
   {journal} {\bibinfo  {journal} {Phys. Rev. Lett.}\ }\textbf {\bibinfo
  {volume} {100}},\ \bibinfo {pages} {100601} (\bibinfo {year}
  {2008})}\BibitemShut {NoStop}%
\bibitem [{\citenamefont {Stolze}\ \emph {et~al.}(1992)\citenamefont {Stolze},
  \citenamefont {Viswanath},\ and\ \citenamefont {M\"uller}}]{Stolze1992}%
  \BibitemOpen
  \bibfield  {author} {\bibinfo {author} {\bibfnamefont {J.}~\bibnamefont
  {Stolze}}, \bibinfo {author} {\bibfnamefont {V.~S.}\ \bibnamefont
  {Viswanath}}, \ and\ \bibinfo {author} {\bibfnamefont {G.}~\bibnamefont
  {M\"uller}},\ }\href {\doibase 10.1007/BF01320828} {\bibfield  {journal}
  {\bibinfo  {journal} {Z. Phys. B. Cond. Mat.}\ }\textbf {\bibinfo {volume}
  {89}},\ \bibinfo {pages} {45} (\bibinfo {year} {1992})}\BibitemShut {NoStop}%
\bibitem [{\citenamefont {Wimmer}(2012)}]{Wimmer2012}%
  \BibitemOpen
  \bibfield  {author} {\bibinfo {author} {\bibfnamefont {M.}~\bibnamefont
  {Wimmer}},\ }\href {\doibase 10.1145/2331130.2331138} {\bibfield  {journal}
  {\bibinfo  {journal} {ACM Trans. Math. Softw.}\ }\textbf {\bibinfo {volume}
  {38}},\ \bibinfo {pages} {30:1} (\bibinfo {year} {2012})}\BibitemShut
  {NoStop}%
\end{thebibliography}%

\end{document}